\documentclass{iopart}
\usepackage{iopams}  
\usepackage{graphicx}
\usepackage{color}
\newcommand{\bm}[1]{\mbox{\boldmath$#1$}}
\newcommand{\be}{\begin{equation}}
\newcommand{\ee}{\end{equation}}
\newcommand{\bea}{\begin{eqnarray}}
\newcommand{\eea}{\end{eqnarray}}
\begin{document}

\title[Exponents of interchain correlation]
{Exponents of interchain correlation for self-avoiding walks  
and knotted self-avoiding polygons}

\author{Erica Uehara and Tetsuo Deguchi}
\address{Department of Physics, Graduate School of Humanities and Sciences,  
Ochanomizu University, 2-1-1 Ohtsuka, Bunkyo-ku, Tokyo 112-8610, Japan}

\eads{\mailto{deguchi@phys.ocha.ac.jp}}

\date{\today}

\begin{abstract} 
We show numerically that critical exponents for two-point 
interchain correlation of an infinite chain 
characterize those of finite chains 
in Self-Avoiding Walk (SAW) and Self-Avoiding Polygon (SAP) 
under a topological constraint. 
We evaluate short-distance exponents $\theta(i,j)$ 
through the probability distribution functions 
of the distance between the $i$th and $j$th vertices of $N$-step SAW 
(or SAP with a knot) for all pairs ($1 \le i, j \le N$).   
We construct the contour plot of $\theta(i,j)$, 
and express it as a function of $i$ and $j$. 
We suggest that it has quite a simple structure. 
Here exponents $\theta(i,j)$ generalize 
des Cloizeaux's three critical exponents 
for short-distance interchain correlation of SAW, 
and we show the crossover among them. 
We also evaluate the diffusion coefficient of knotted SAP 
for a few knot types, which can be calculated 
with the probability distribution functions 
of the distance between two nodes. 
\end{abstract}

\maketitle

\section{Introduction}
Polymers with nontrivial topology such as cyclic polymers 
have attracted much interest in several fields. 
Ring polymers are observed in nature such as circular DNA 
whose topology is given by the trivial knot (Fig. 1),  
while DNA with nontrivial knots are derived in experiments 
\cite{Nature-trefoil,DNAKnots,Bates}. 
Naturally occurring proteins whose ends connected to give a circular topology 
has been recently discovered \cite{Craik}. 
Due to novel developments in experimental techniques, 
ring polymers are now effectively synthesized in chemistry 
 \cite{Grubbs,Takano05,Takano07,Grayson,Tezuka2010}.  
Moreover, polymers of topologically complex structures, 
which are sometimes called topological polymers, 
have been synthesized and separated 
with respect to their hydrodynamic radii such as through GPC 
\cite{Tezuka2011,Tezuka-book}. 
It is thus an interesting theoretical problem to calculate 
physical quantities such as the hydrodynamic radius, i.e. 
the diffusion coefficient,  of each topological type.  
It can be derived through interchain correlation of the polymer chain 
by Kirkwood's approximation. 
Here we remark that the topology of a ring polymer is specified by a knot, and 
it gives the simplest and most fundamental example of nontrivial topologies. 

Topological constraints often play a central role in the statistical and dynamical properties of ring polymers in solution 
\cite{Whitt07,Micheletti,knot2010}. 
For instance, the mean-square radius of gyration of ring polymers  
under a topological constraint can be much larger than 
that of no topological constraint, in particular, 
at the $\theta$ temperature of the corresponding linear polymers 
\cite{Deutsch,GrosbergPRL,SAP02,JPhysA-Lett,Akos03,Matsuda,Moore}. 
We call the phenomenon {\it topological swelling}. 
It is also confirmed in an experiment \cite{Takano-Macromolecules}. 
Due to the strong finite-size effect, however, 
it is not easy to determine numerically the exponent of the mean square radius 
of gyration for knotted ring polymers in $\theta$ solution 
\cite{Matsuda}. 
It is thus interesting to study topological effects on the scaling behavior 
of two-point correlations for Self-Avoiding Polygons or random polygons  
through simulation.

The Self-Avoiding Walk (SAW) and the Self-Avoiding Polygon (SAP) are 
fundamental theoretical models for linear and ring polymers in good solution, respectively \cite{deGennes,Doi-Edwards}. 
The scaling behavior of SAW is studied through the Monte-Carlo simulation 
and the renormalization group approach \cite{Zinn-Justin}.   
Correlations among configurations of SAW and SAP are nontrivial 
due to the excluded volume effect, and have attracted much interest 
in theoretical studies \cite{Fisher,McKenzie,desCloizeaux1974,desCloizeaux,Baumgartner,Oono,Geometrical,Swelling,Bishop,Valleau,Timo,Vicari}. 
SAP has several different points from SAW: 
It is not only that SAP has cyclic symmetry while
SAW has two ends and no translational symmetry, 
but also that SAP has a topological constraint specified by a knot. 
However, we shall show that several short-chain scaling properties of SAW 
are useful for describing those of SAP. 

\begin{figure}[htpb] 
\begin{center}
\includegraphics[width=5cm,clip]{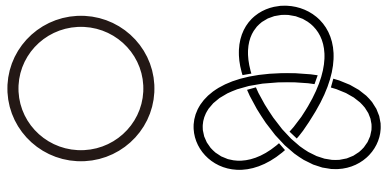}
\caption{Trivial knot, $0_1$ (left), and the trefoil knot, $3_1$ (right).  }
\label{fig:knot}
\end{center} 
\end{figure}

In this paper, 
we study the scaling behavior of interchain correlation 
for SAW and off-lattice SAP with a fixed knot type 
through the Monte-Carlo simulation. In particular, we evaluate 
exponents for short-distance correlation for any pair of segments 
in a SAW or SAP under a topological constarint.  
We numerically determine the probability distribution function of the distance between two vertices $i$ and $j$ of the chain, 
from which we evaluate the exponent $\theta(i,j)$ for 
short-distance correlation and the exponent $\delta$ 
for long-distance asymptotic behavior.   
The estimates of exponents $\theta(i,j)$ and $\delta$ 
are useful for expressing physical quantities such as 
the diffusion coefficient and the structure factor of SAW or SAP 
as an approximate integral form. 
For SAW, exponents $\theta(i,j)$ generalize 
des Cloizeaux's three exponents $\theta_s$ ($s=0,1,2)$ 
for short-distance interchain correlation of an infinite chain,  
which we shall define shortly. 
We shall show that the estimates of exponents $\theta(i,j)$ corresponding to 
$\theta_s$ ($s=0,1,2)$ are roughly similar to or a little smaller than 
the theoretical values of $\theta_s$. 
The difference may be due to the finiteness of the chain investigated.  
We also show the crossover among them. 
For SAP consisting of cylindrical segments 
we show that exponents $\theta(i,j)$  and $\delta$  
of SAP with large excluded volume 
are close to those of SAW, while exponents $\theta(i,j)$  and $\delta$  for 
SAP with small excluded volume are much smaller than those of SAW 
and close to those of Random Walks.

Let us briefly review the scaling behavior of interchain correlation of SAW.  
We denote by $p_0({\bm r}; N)$  the probability distribution function 
of the end-to-end vector ${\bm r}$ of an $N$-step SAW. Considering the 
rotational symmetry we express it also as $p_0(r; N)$ where $r$ is the 
end-to-end distance: $r=|{\bm r}|$.  
The large-$r$ asymptotic behavior of $p_0(r; N)$ 
was argued \cite{Fisher} as   
\be 
p_0(r; N) \sim R_N^{-d} A(r/R_N) \exp\left(- (r/R_N)^{\delta} \right) 
\ee
where $R_N= R_0 N^{\nu}$ and $\delta=1/(1-\nu)$.  
Here the scaling exponent $\nu$ is given by $\nu \approx 0.588$.
The small-$r$ behavior of $p_0(r; N)$ 
was studied analytically \cite{McKenzie}. 
By assuming a scaling function $F_0(y)$ satisfying 
$p_0(r; N) = R_N^{-d} F_0(r/R_N)$, 
it was shown that the short distance behavior is given by  
\be 
F_0(y) \sim y^{g}  \quad  {\mbox as}  \quad  y \rightarrow 0 \, . 
 \label{eq:F_0}
\ee 
with $g=(\gamma+ 1 - d \nu - \alpha)/\nu$.  
The exponent $g$ was also derived 
through renormalization group (RG) arguments \cite{desCloizeaux1974}. 
Here we have $g=(\gamma-1)/\nu$ through the scaling relation: 
$\alpha=2 -d \nu$. It is also derived via RG arguments  
with the blob picture \cite{deGennes}.

Short-distance correlation between two points of a long polymer 
in a good solvent was studied by des Cloizeaux 
with the RG techniques \cite{desCloizeaux}. 
The exponents of the short-distance correlation,  
$\theta_s$ for $s=0, 1, 2$, were defined 
through the probability distribution functions $p_s(r, N)$ 
of the distance between two given vertices 
of an $N$-step SAW in the large-$N$ limit as follows. 
We denote by $p_1(r, N)$ the probability distribution 
function of the distance between an end point and 
a middle point of the SAW, and by $p_2(r, N)$ that of 
the distance between two points in the middle region of the SAW.  
Assuming that $p_s(r, N) \approx  R_N^{-d} F_s(r/R_N)$, 
we define the critical exponents $\theta_s$ 
for short distance correlation by  
\be 
F_s(y) \sim y^{\theta_s} \quad {\mbox as}  \quad y \rightarrow 0 \,,    
 \quad \mbox{for} \quad s =0, 1, 2.  
\ee
Here we remark that exponent $g$ in eq. (\ref{eq:F_0}) corresponds to 
$\theta_0$.  The exponents $\theta_s$ for $s=0, 1, 2$ 
were calculated by des Cloizeaux 
with RG techniques in terms of the $\epsilon$-expansion upto the second order  
\cite{desCloizeaux}. 
The estimates for $d=3$ are given by 
\be 
\theta_0^{(RG)} = 0.273, \quad \theta_1^{(RG)}=0.46, 
\quad \theta_2^{(RG)}= 0.71 .    
\label{eq:theta-values}
\ee
In the RG derivation of $\theta_s$ 
it is assumed that the SAW is infinitely long. 
Here remark that making use of the estimates of critical exponents 
$\gamma$ and $\nu$ of the $O(N)$ model with higher loop corrections,   
 we evaluate $\theta_0$ through the relation $\theta_0=g=(\gamma-1)/\nu$ 
as follows. We have 
$\theta_0= 0.2713 \pm 0.0039$ from the estimates (d=3 expansion) 
in Fig. 1 of Ref. \cite{Zinn-Justin}, 
and $\theta_0= 0.2672 \pm 0.0056$ from 
the estimates ($\epsilon$-expansion, bc) 
in Fig. 2 of Ref. \cite{Zinn-Justin}.

As a first result of the present paper, we numerically evaluate the critical 
exponents $\theta_s$ for $s=0, 1, 2$, by evaluating 
$\theta(i,j)$ in the simulation 
of $N$-step SAW on the cubic lattice with $N=8000$,  
and compare them with the theoretical values obtained 
by des Cloizeaux. Moreover, we show the crossover among exponents 
from $\theta_1$ and $\theta_2$ to $\theta_0$, 
and that between $\theta_1$ and $\theta_2$.  
Let us denote the estimates of $\theta_s$ ($s=0,1, 2$) 
by $\theta_s^{(MC)}$ ($s=0,1, 2$). They are given by 
\be 
\theta_0^{(MC)} = 0.23 \pm 0.02, \, 
\theta_1^{(MC)} = 0.35 \pm 0.03, \, 
\theta_2^{(MC)} = 0.74 \pm 0.03 \, . 
\label{eq:theta-MC}
\ee
The estimate of $s=0$ is roughly 
the same but a little smaller than the RG value 
with respect to errors. It may be due to the finiteness of the chain. 
The estimate of $s=1$ is clearly smaller than the RG value; 
The estimate of $s=2$ is roughly the same with 
the theoretical value within errors. 
For SAP with knot $K$ the estimates of the exponent 
for short-distance correlation, denoted by $\theta_K$ or $\theta_K(\lambda)$, 
are a little  smaller than the RG value of $\theta_2$ for SAW.  
We have $\theta_K=0.679 \pm 0.004$ for SAP of the trivial knot $0_1$ 
 (i.e., $K=0_1$) 
with $N=3000$ between two nodes separated by $900$ steps along the chain 
(i.e., $\lambda=0.3$).    

The estimates of critical exponents of a finite chain are useful, 
although the simulation values may be slightly different 
from the theoretical values due to the finiteness of the chain. 
For instance, with the estimates $\theta^{(MC)}(i,j)$ 
for exponents $\theta(i,j)$ we have good fitting curves 
to the probability distribution function of the distance 
between two verstices $i$ and $j$ of SAW. 
We can thus approximate it by an analytic function in terms of 
$\theta^{(MC)}(i,j)$. Moreover, polymers in reality are always of finite length, which can be compared with simulation results of finite chains.

For SAW the end-to-end distance distribution \cite{Bishop,Valleau} 
and the probability distribution functions 
of the internal distances \cite{Baumgartner,Timo} 
have been evaluated numerically in simulation. 
However, the critical exponent of the short-distance correlation 
have not been evaluated in simulation with high numerical precision, yet. 

The present study  of the two-point correlations of SAW is 
also important in expressing topological effects of knotted ring polymers 
in $\theta$ solution 
\cite{desCloizeaux-Let,distance,scattering,OCAMI,Akita,JiroSuzuki2013}. 
It has been shown that the average size of a ring polymer with a fixed knot in $\theta$ solution becomes enhanced due to the topological entropic force acting 
among the segments of the ring polymer. 
Furthermore, it has been suggested 
in several researches that the universality class of 
ring polymers in $\theta$ solution should be given by that of SAW 
\cite{GrosbergPRL,Akos03,Matsuda,Moore}. 
However, the distribution function of the distance 
between two vertices of a random polygon with fixed knot type is close 
to the Gaussian one \cite{distance,Akita}. 
Thus, the method of the present paper for determining the exponents of 
two-point correlation of SAW or SAP  is useful for investigating 
the critical behaviour of ring polymers with a fixed knot in $\theta$ solution 
more explicitly.   

The contents of the paper consist of the following. 
In \S 2, we explain the numerical methods of simulation in this research. 
We introduce normalized distance $x$, which is given by the 
distance between two points of SAW divided 
by the square root of the mean-square distance. 
We then define the probability 
distribution function of variable $x$. 
In \S 3, we define exponents $\theta(i,j)$ between the $i$th and 
$j$th vertices of SAW, and  give the numerical results. 
We show that the fitting formula for the probability distribution function 
of the distance between two vertices of SAP as a function of 
normalized distance $x$ gives good fitting curves to the numerical data. 
We then derive numerical estimates of exponents $\theta(i,j)$ 
for all pairs of vertices $i$ and $j$, and express them 
as a function of $i$ and $j$. 
In particular, we present the contour plot of exponents 
$\theta(i,j)$ for all vertices  $i$ and $j$ ($0 \le i, j \le N$), 
and show that it has a simple structure.   
In \S 4 we briefly review the  
topological swelling, i.e., the enhancement of the mean square radius of gyration due to topological constraints. Then, 
we show the scaling behaviour of interchain correlation 
of SAP under a given topological constraint. 
In \S 5 we evaluate the diffusion coefficient of 
SAP with a fixed knot through Kirkwood's approximation 
for a few knots.  In \S 6 we give concluding remarks.

%
%
\section{Numerical methods and important notation}

%
%
\subsection{Pivot algorithm for generating Self-Avoiding Walks}

We have generated $10^5$ configurations of the $N$-step SAW 
on the cubic lattice by the pivot algorithm \cite{Pivot,Madras-Orlitsky-Shepp} 
for several values of $N$ with $N \le 8000$. 
For a given initial configuration of the $N$-step SAW 
we pick up the configuration of SAW after every $8N$ Monte-Carlo procedures, 
and assume that it is independent from the previous one. 
Here, in the cubic lattice each edge has unit length.

\begin{figure}[bht] 
\begin{center}
\includegraphics[width=8cm,clip]{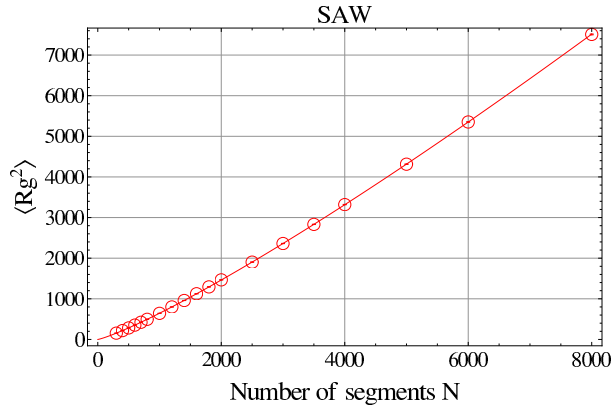}
\caption{Mean-square radius of gyration for SAW with $N$ steps  
on the cubic lattice. 
The best estimates of eq. (\ref{eq:fit-Rg})  
are given by 
$\nu= 0.5899 \pm 0.0004$, $A=0.187 \pm 0.001$ and $B=-3.0 \pm 0.9$. 
The $\chi^2$-value per datum is given by 0.94. } 
\label{fig:Rg-SAW} \end{center} \end{figure}

The mean-square radius of gyration, $R_g^2$, 
for SAW on the cubic lattice is evaluated  
for several different numbers of $N$. 
The fitting curve in Fig. \ref{fig:Rg-SAW} is given by  
\be 
R_g^2 = A N^{2 \nu} (1 + B/N ) \, .  \label{eq:fit-Rg} 
\ee
Here the estimate of exponent $\nu$ is given by 
$\nu= 0.5899 \pm 0.0004$, which is consistent 
with exponent $\nu_{\rm SAW}$ of SAW.

%
%
\subsection{Distance between two vertices of SAW}

\begin{figure}[htpb] 
\begin{center}
\includegraphics[width=6cm,clip]{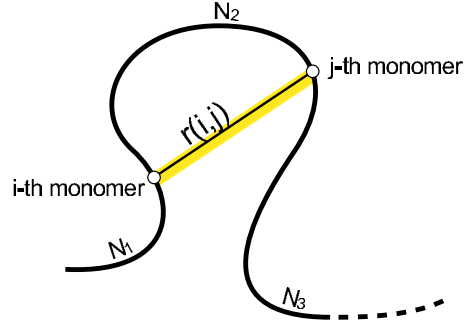}
\caption{Distance $r(i,j)$ between the $i$th and $j$th vertices 
of a long Self-Avoiding Walk (SAW) of $N$ steps, where $1 \le i < j \le N$ and $N \gg1$. We have $p_1(r, N)$ for $N_1=0$ and $N_2 \approx N/2$, and $p_2(r, N)$  for $N_1, N_2, N_3 \gg 1$  \cite{desCloizeaux}.  
We define parameter $\lambda$ by $N_2=\lambda N$ and parameter $\mu$ by $N_1= \mu \lambda N$.} 
\label{fig:SAW}
\end{center} 
\end{figure}

For a given three-dimensional 
configuration of SAW with $N$ steps on the cubic lattice, 
we take a pair of integers $i$ and $j$ such that 
$0 \le i < j \le N$ (see, Fig. \ref{fig:SAW}). 
There are $N+1$ vertices in total 
from the 0th to the $N$th vertex 
with position vectors ${\vec R}_j$ for $j=0, 1, \ldots, N$.  
To the given configuration of SAW 
we calculate the distance between the $i$th  and $j$th vertices   
\be 
r(i, j) = | {\vec R}_{j} - {\vec R}_{i}|. 
\ee
We shall denote $r(i,j)$ also by $r_{i, j}$, briefly.  
%

For integers $i$ and $j$ satisfying $1 \le i < j \le N$,  
let $N_1$, $N_2$ and $N_3$ denote the number of steps in the first, second and third  part of an $N$-step SAW, respectively, as shown in Fig. \ref{fig:SAW}.  
We have $N_1=i$, $N_2=j-i$, and  $N_3=N-j$.

We denote by $R_N(i,j)$  
the average distance between the $i$th 
and $j$th vertices of an N-step SAW, i.e. 
the square root of the mean square distance 
between the $i$th and $j$th vertices of SAW with $N$ steps:    
$R_{N}(i,j) = \sqrt{\langle r^2_{ij} \rangle}$. 
It is approximated by 
\be 
\left( R_N(i,j) \right)^2 = A_{i,j} |j-i|^{2 \nu} 
\left(1 + B_{i,j}/|j-i| \right)  \, . \label{eq:Rij}
\ee
Here $A_{i,j}$ and $B_{i,j}$ are fitting parameters.

We introduce parameter $\lambda$ such that selected two points $i$ and $j$ 
are separated by $\lambda N$ steps ($0 \le \lambda \le 1$): 
$N_2 = \lambda N$.   
We define parameter $\mu$ by $N_1= \mu N_2$. We have 
\be 
N_1 = \mu N_2 = \mu \lambda N \, . 
\ee
In terms of parameters $\lambda$ and $\mu$, integers $i$ and $j$ are given by  
$i=N_1= \mu \lambda N$ and $j= N_1 + N_2 = (1+\mu) \lambda N$, respectively. 
The intger $N_3$ is givben by $N_3=N-j= N - (1+\mu) \lambda N$. 
In the case of $s=2$ where $N_1= N_3$,  
we have $(1+ 2\mu) \lambda =1$.

\begin{figure}[htb] \begin{center}
\includegraphics[width=8cm,clip]{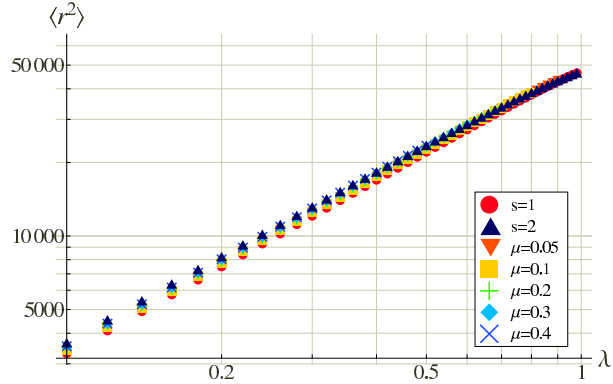}
\caption{The mean-square distance, $R_N^2(i,j) = \langle r^2_{i,j} \rangle$, between two points $i$ and $j$ 
separated by $\lambda N$ steps in a SAW of $N$ steps on the cubic lattice, 
where  $j-i= \lambda N$ and $i= \mu \lambda N$ for $\mu$ = 0, 0.05, 0.1, 0.2, 0.3, and 0.4. Here $N=8000$.  The case of $s=1$ corresponds to $\mu=0$.  
} 
\label{fig:Two-point-distance_SAW}
\end{center} 
\end{figure}

The data of the mean square distance  $\langle r^2_{i,j} \rangle$ 
versus $\lambda$ are shown in Fig. \ref{fig:Two-point-distance_SAW}.  
The data points of different values of parameter $\mu$ 
almost coincide each other for $0 \le \mu \le 0.5$ 
in Fig. \ref{fig:Two-point-distance_SAW}. 
It is thus suggested that the fitting parameters 
of (\ref{eq:Rij}) are 
continuous with respect to the change of parameter $\mu$.

\begin{table}[htbp] \begin{center} 
\begin{tabular}{c|c|c|c}  
 & $\nu$  & $A_{i,j}$  & $B_{i,j}$ \\ \hline 
$s=1$ & 0.589 $\pm 0.0007$  &  1.26 $\pm 0.02$  & -3 $\pm 2$    \\ \hline 
$\mu=0.05$ & 0.5888 $\pm 0.0009$  &  1.29 $\pm 0.02$  & $-3 \pm 3$  \\ \hline 
$\mu=0.1$ & 0.5877 $\pm 0.0008$  &  1.33 $\pm 0.02$  & $-6 \pm 3$  \\ \hline 
$\mu=0.2$ & 0.5881 $\pm 0.001$  &  1.34 $\pm 0.02$  & $-5 \pm 3$  \\ \hline 
$\mu=0.3$ & 0.589 $\pm 0.001$  &  1.34 $\pm 0.03$  & $-2 \pm 4$  \\ \hline 
$\mu=0.4$ & 0.585 $\pm 0.003$  &  1.45 $\pm 0.06$   & $-16 \pm 8$  \\ \hline  
$s=2$ & 0.562 $\pm 0.001$ & 2.13 $\pm 0.05$  & -52 $\pm 4$    \\ \hline 
 \end{tabular} 
\caption{Estimates of parameters for the graph of 
the mean-square diatance of two points of SAW separated by $\lambda N$ steps 
versus parameter $\lambda$: $\langle r^2(i,j) \rangle= 
A_{i,j} (\lambda N)^{2 \nu} (1 + B_{i,j}/(\lambda N))$ 
where $j-i=\lambda N$, $i=\mu \lambda N$ and $0.1 \le \lambda \le 0.5$. 
Here $N=8,000$.} \label{table:main} 
\end{center} \end{table}

We consider two special types of interchain distance:  
$R_{N, s}(\lambda)$ for $s=1, 2$. 
We define $R_{N, 1}(\lambda)$ 
by the square root of the mean-square distance 
between the 0th and $\lambda N$th vertices. Here we recall $i=0$ 
and $j=\lambda N$: 
$R_{N, 1}(\lambda) = \sqrt{ \left\langle r^2(0, \lambda N) \right\rangle}$.   
We define $R_{N, 2}(\lambda)$ by  
the square root of the mean-square distance 
between the $(1 \pm \lambda)N/2$ th vertices. Here, we have  
$i=(1 - \lambda)N/2$ and $j=(1 + \lambda)N/2$, and parameter $\mu$ is given by 
$\mu=(1-\lambda)/(2\lambda)$.  Thus, we have 
$R_{N, 2}(\lambda) = 
\sqrt{ \left\langle r^2((1-\lambda) N/2, (1+\lambda) N/2) \right\rangle}$. 
The square roots of interchain distances $R_{N, s}(\lambda)$ for $s=1, 2$  
are well approximated by 
\be 
\left( R_{N,s}(\lambda) \right)^2 = A_s (\lambda N)^{2 \nu} 
\left(1 + B_s /(\lambda N) \right),  
\ee   
where $A_s$ and $B_s$ are fitting parameters. 
We shall often neglect the correction term $B_s/(\lambda N)$.

%
%
\subsection{Probability distribution function of the distance between two vertices of SAW}

Let us denote by $p({\bm r}; i, j; N)d^3 {\bm r}$ 
the probability of finding the $j$th vertex 
in the region $d^3 {\bm r}$ at a position 
${\bm r}$ from the $i$th vertex of an $N$-step SAW. 
It is expressed in terms of the average 
over all configurations of SAW, $\langle \cdots \rangle$, as follows.   
\be 
p({\bm r}; i, j; N) = 
\langle \delta({\bm r} - ({\bm R}_j- {\bm R}_i)) \rangle .  
\ee
Here we recall that ${\bm R}_i$ and ${\bm R}_j$ are the position vectors of 
the $i$th and $j$th vertices of the SAW, respectively.  
Due to the rotational symmetry,  the probability distribution function 
$p({\bm r}; i, j; N)$ depends only on the distance $r=| {\bm r}|$,  
and we denote it simply by $p(r; i,j; N)$. 

We shall show that good fitting curves are given by 
the following formula 
\be 
p(r; i, j ; N) = c_{i,j} \,  \left(r/R_{N}(i,j) \right)^{\theta(i,j)} 
\exp \left( - ( D_{i,j} \, r/ R_{N}(i,j) )^{\delta} \right) \, . 
\label{eq:universal-p(i,j)}
\ee
Here exponent $\delta$ 
is related to the exponent $\nu$ by $\delta=1/({1-\nu})$, and 
$R_{N}(i,j)$ are given by the square root of the mean-square distance 
between the two vertices $i$ and $j$. 
Applying it to the numerical data of $p(r; i,j; N)$, 
we evaluate exponents $\theta(i,j)$ 
by the best estimates of parameters for fitting curves.

Let us consider two special types of the probability distribution functions 
of the distance $r$ between two vertices of an $N$-step SAW:  
$p_s(r; \lambda, N)$ for $s=1, 2$.  
For $s=1$,  $p_1(r; \lambda, N)$ is defined for the distance $r$ 
between the vertex of an end point and another vertex of SAW, 
say, the $n$th vertex with $n=\lambda N$; 
for $s=2$, $p_2(r; \lambda, N)$ is defined for the distance $r$ 
between the $(1 - \lambda)N/2$th and $(1 + \lambda)N/2$th vertices of SAW.  
Here we recall that for $s=0$, the probability distribution function 
$p_0(r; N)$ has been defined for the distance $r$ 
between two ends of an $N$-step SAW. 
It corresponds to  $p_s(r; \lambda, N)$ 
for $s=1, 2$ in the case of $\lambda=1$.

We shall also show that $p_{s}(r; \lambda, N)$ 
for $s=1, 2$ are well approximated by 
\be p_{s}(r; \lambda, N) = c_s \,  \left(r/R_{N, s}(\lambda) 
\right)^{\theta_s(\lambda)} 
\exp \left( - (D_s \, r/R_{N, s}(\lambda))^{\delta} \right) \, . 
\label{eq:universal-p}
\ee 
We shall evaluate exponents $\theta_s(\lambda)$ for $s=1, 2$ by 
the fitting formula of $p_s(r; \lambda, N)$ for $s=1, 2$, respectively.

%
%
\subsection{Distribution function of the normalized distance}

Let us assume an ensemble of SAW where there are $W=10^5$
 random configurations 
of $N$-step SAW on the cubic lattice. 
For the distance between the $i$th and $j$th vertices, $r_{i, j}$,  
we introduce the normalized distance $x_{i, j}$ by 
\begin{equation} 
x_{i, j} = r_{i, j} / \sqrt{\langle r_{i, j}^2 \rangle} . 
\end{equation}
We set the length $\Delta x$ of intervals   
by $\Delta x=10^{-1}$. 
We enumerate the number of configurations 
of  SAW such that the normalized distance $x_{i, j}$ 
 between the $i$th and $j$th vertices 
satisfies the conditions $x < x_{ij} < x + \Delta x$.    
We express the number by $n_{ij}(x, \Delta x)$. 
We define the probability distribution function 
$f(x; i, j; N)$ of the normalized distance $x$ 
between the $i$th and $j$th vertices by   
\begin{equation} 
x^2 f(x; i, j; N) \Delta x = n_{ij}(x, \Delta x)/W  \, . 
\end{equation}
In terms of $p(r; i,j ; N)$ we have 
$f(x; i, j; N) = 4 \pi R^3_N(i,j) \, p(r; i, j; N)$.  
Hereafter we also call $f(x; i, j; N)$ distribution function.  

Let us now introduce symbols $f_s(x; \lambda, N)$ for $s=0, 1, 2$.  
We denote by $f_0(x; N)$ 
the probability distribution function of the normalized 
end-to-end distance $x=r/R_N$.     
We then denote by $f_{1}(x; \lambda, N)$ the probability 
distribution function of the normalized distance 
between an end point (the 0th vertex) 
and the $\lambda N$th vertex of SAW of $N$ steps and  
by $f_{2}(x; \lambda, N)$ that 
of the normalized distance between two vertices separated by $\lambda N$ steps 
in a middle region of SAW.

%
%
\subsection{Algorithm for constructing off-lattice SAP}

We generated $2 \times 10^5$ configurations of SAP consisting of $N$ cylindrical segments with cylindrical radius $r_{ex}$ of unit length for various 
number of nodes $N$.  
Each cylinder segment has the excluded volume of $\pi r_{ex}^2$. 
In the model of cylindrical SAP, we assume that neighboring segments 
have no excluded volume interaction: 
Neighboring cylinder segments may overlap each other. 

In the Monte-Carlo procedure 
we  first select two nodes of SAP randomly and consider a subchain 
between the two nodes. We then constructed the ensembles of cylindrical SAP by 
combinig the crank-shaft move 
and the rotation of a subchain 
of cylindrical SAP around an axis at the center of the axis by 180 degrees. 
Here, the axis is orthogonal to the end-to-end vector of the subchain 
We apply the crank-shaft move $2N$ times, 
and then we apply the rotation of subchains $2N$ times 
in the Monte-Carlo algorithm.

%
%
%
\section{Scaling behavior of interchain correlation of SAW}

%
%
\subsection{Distribution functions $f_s(x; \lambda; N)$ of the distance 
between two vertices of SAW and short-distance exponents $\theta_s$}

Let us introduce the formula for fitting curves to the data of   
the probability distribution functions $f_s(x; \lambda, N)$ for $s=1, 2$ 
as follows.  
\be 
f_s(x; \lambda, N) = C_{s} \, x^{\theta_s(\lambda)} 
\exp \left( - 
(D_{s} \, x)^{\delta} \right) \,  \label{eq:universal-f}
\ee
where $\delta= 1/(1-\nu)$. The constants $D_s$ and $C_s$ are given by 
\bea 
D_s & = & \sqrt{\frac {\Gamma((5+ \theta_s)/\delta) }
                       {\Gamma((3+ \theta_s)/\delta) } } \, ,  \nonumber \\ 
C_s &  = &  {\frac {\delta} {\Gamma((3+\theta_s)/\delta)}} 
\left(  {\frac {\Gamma((5+ \theta_s)/\delta) }
                       {\Gamma((3+ \theta_s)/\delta) }} 
  \right)^{(3+\theta_s)/\delta} \, . \label{eq:CD} 
\eea
Here we recall that $x$ denotes the normalized distance: 
$x = r/R_{N,s}(\lambda)$, where $R_{N,s}(\lambda) = \sqrt{A_s} \, 
(\lambda N)^{\nu}$. 
For $s=0$, we assume that $R_{N,s}(\lambda)$ denotes the end-to-end distance 
$R_N$, and apply the formula 
which is obtained by replacing  all $\theta_s(\lambda)$ of  
(\ref{eq:universal-f}) and (\ref{eq:CD}) with $\theta_0$.

Formula  (\ref{eq:universal-f}) has two fitting parameters 
$\theta_s(\lambda)$ and $\delta$ for each $s$ of $s=1, 2$. 
The constants $C_s$ and $D_s$ satisfy 
the followng constraints for $s=1, 2$: 
\be 
 \int_0^{\infty} x^2 f_s(x; \lambda, N)  dx = 1 \, , \quad 
\int_0^{\infty} x^4 f_s(x; \lambda, N) dx = 1 \, .   
\ee

We made the graphs of the distribution function of 
the end-to-end distance $f_0(x; N)$ ($s= 0$)
and those of the distribution functions 
$f_s(x; \lambda, N)$  of the distance 
between two vertices separated by $\lambda N$ steps for $s=1, 2$ 
with 45 different vaues of $\lambda$ from 0.10 to 0.98 by 0.02 
 against normalized distance $x$. Each graph has 20 data points from $x=0.05$ 
to $1.95$. Here we recall that $10^5$ SAWs of $N$ steps are 
generated by the pivot algorithm for $N=8,000$.

\begin{figure}[htbp]
\begin{center}
\includegraphics[width=9cm,clip]{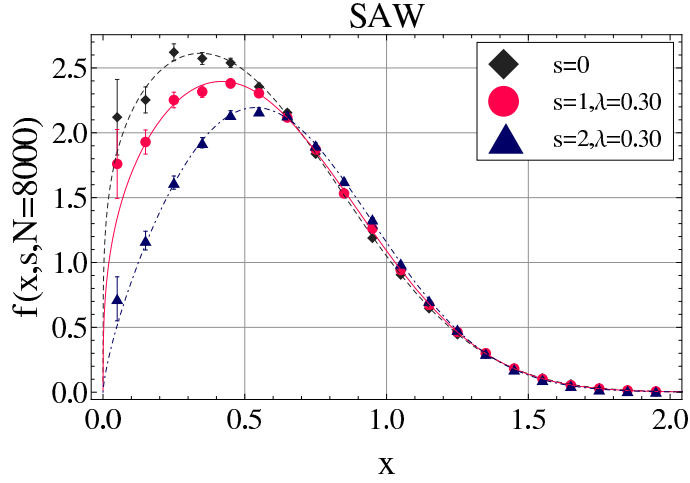}
\caption{Distribution function $f_s(x; \lambda, N)$ 
of the distance between two vertices 
of Self-Avoiding Walk of $N$ steps separated by $\lambda N$ steps 
for $s=1, 2$, with $\lambda= 0.3$ and $N=8,000$.
Distribution function of the end-to-end distance, 
$f_0(x; N)$, is also plotted.  
Each distribution function has 20 data-points. 
}
\label{fig:linear}
\end{center}
\end{figure}

Formula (\ref{eq:universal-f}) gives good fitting curves 
to the data of the probability distribution functions $f_s(x; \lambda, N)$ 
of the distance between two vertices of SAW for several different values of $\lambda$ and $N$ and over almost the entire region of normalized distance $x$. 
The $\chi^2$ value per datum is less than 2.0 for all fitting curves 
(in total, 91 curves).

We have thus shown that the probability distribution function 
of the distance between two vertices of type $s$ ($s=0,1,2$) 
is given by 
\be 
p_s(r; \lambda, N) =  
f_s(r/R_{N,s}(\lambda); \lambda, N)/ 
\left( 4 \pi R^3_{N,s}(\lambda) \right) \, .  
\ee
where $R_{N,s}(\lambda) = \sqrt{A_s} (\lambda N)^{\nu}$ for $s=1, 2$.

We have evaluated parameters $\delta(\lambda)$ (or $\nu(\lambda)$) and 
$\theta_s(\lambda)$ for $s=1,2$ with the least-square method 
by applying formula (\ref{eq:universal-f}) 
to the data plots of the probability distribution functions 
$f_s(x; \lambda, N)$ for $s=1,2$ as functions of normalized distance $x$ 
over the entire region of $x$ for various  values of $\lambda$ 
$(0.1 < \lambda < 1.0)$. Here, $N=8,000$. 
For a given value of $\lambda$ and each of  $s=1, 2$, 
we make a fitting curve to the data points of 
distribution function $f_s(x; \lambda, N)$ of the distance 
between two vertices, and evaluate fitting parameters 
$\delta(\lambda)$ and $\theta_s(\lambda)$.

\begin{figure}[htpb] 
\begin{center}
\includegraphics[width=9cm,clip]{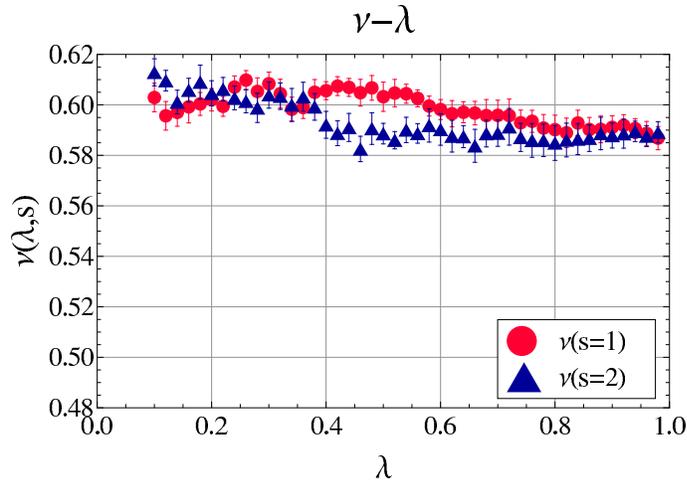}
\caption{Exponents $\nu(\lambda)$ evaluated through the distribution functions 
$f_s(x; \lambda, N)$ 
of the distance between two points separated by $\lambda N$ steps of 
SAW with $N=8,000$ for $0.1 \le \lambda < 1$.  
Here, $\nu$ is calcuted from $\delta$ by $\nu=1-1/\delta$.} 
\label{fig:exponent-nu-SAW}
\end{center} 
\end{figure}

The estimates of $\nu(\lambda)$ evauated from $\delta(\lambda)$ 
in formula (\ref{eq:universal-f}) of distribution functions $f_s(x; \lambda, N)$ for $s=1,2$  via relation $\nu=1-1/\delta$,
 are plotted against parameter $\lambda$ in Fig. \ref{fig:exponent-nu-SAW}.  
They are almost completely constant with respect to $\lambda$, 
and consistent with the exponent of SAW, $\nu_{\rm SAW} \approx 0.588$.

For an illustration, we presented in Fig. \ref{fig:linear} 
fitting curves to distribution functions 
$f_s(x; \lambda, N)$ for $s=1$ and 2, respectively. 
In the cases of $s=1$ and 2 
the curves for the data-points of $\lambda=0.30$ 
 coincide within errors for any value of $x$.  

\begin{table}[htbp] \begin{center}
\begin{tabular}{c|cc|cc|cc}   \hline   
& $s=0$ &   & $s=1$ &  &  $s=2$ &   \\  
& &      &  $\lambda=0.3$ & &  $\lambda=0.3$ &  \\ \hline 
 $\theta_s$ &  
0.23 & $\pm 0.02$ & 0.33 & $\pm 0.03$ & 0.73 $\pm 0.03$    \\ \hline 
  $\nu$ & 0.589 &  $\pm 0.005$ 
& 0.608 & $\pm 0.005$ & 0.604 & $\pm 0.005$ \\ \hline 
 $\chi^2$/datum & 0.786  &  & 0.887 &  & 1.07 &  \\ \hline 
 \end{tabular} 
\caption{Estimates of fitting parameters and the $\chi^2$ value per datum 
for the fitting curves in Fig. \ref{fig:linear}: 
$\theta_s(\lambda)$ and $\delta$ for $s=1, 2$. 
 Here $N=8,000$ and $\lambda=$ 0.3.  }
\label{table:main} 
\end{center}
\end{table}

The exponents $\theta_s$ for $s=0,1,2$ 
are in increasing order: $\theta_0 < \theta_1 < \theta_2$.  
We observe in Fig. \ref{fig:linear} 
that in small $x$ region, 
the fitting curve of $s=0$ is higher in position than $s=1$, and 
the fitting curve of $s=1$ is higher in position than 
the  fitting curve of $s=2$. 
%
%
In fact, by taking the derivative of fitting formula  
(\ref{eq:universal-f}), we can show that 
the peak position of the fitting curve (\ref{eq:universal-f}) 
becomes larger as parameter $\theta_s$ increases.

\begin{figure}[htpb] 
\begin{center}
\includegraphics[width=10cm,clip]{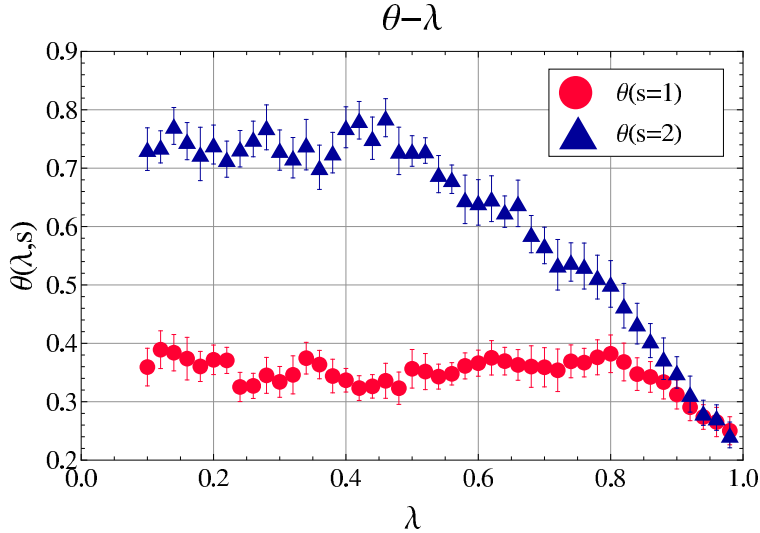} 
\caption{
exponents $\theta_1(\lambda)$ 
and $\theta_2(\lambda)$ for  
 $0.1 \le \lambda < 1$ and $N=8,000$.} 
\label{fig:exponent8T}
\end{center} 
\end{figure}

The critical exponents $\theta_1(\lambda)$ and $\theta_2(\lambda)$ 
are plotted against parameter $\lambda$ 
over a wide range such as  $0.1 \le \lambda < 1.0$ 
in Fig. \ref{fig:exponent8T} for SAW of $N=8, 000$ steps. 
They are given by the best estimates that are obtained by applying 
formula (\ref{eq:universal-f}) 
to the data. Here we recall that formula (\ref{eq:universal-f}) has only two 
fitting parameters, $\theta$ and $\delta$.  

We observe that the estimates of $\theta_1(\lambda)$ 
are independent of parameter $\lambda$ for $0.1 < \lambda < 0.8$.   
The constant value of $\theta_1(\lambda)$ is given by 0.35 
(see also eq. (\ref{eq:theta-MC}9), 
which is a little smaller than the theoretical value: $\theta_1=0.46$. 
Here we remark that in the theoretical derivation \cite{desCloizeaux} 
the remaining part of the chain is assumed to be infinitely long; i.e., $N_3 \rightarrow \infty$. However, when $\lambda < 0.8$, 
the remaining part of SAW is more than 20 percentage of the SAW, 
which may be long enough in the case of $N=8, 000$.  
We also observe that for $0.1 < \lambda < 0.5$, 
the estimates of $\theta_2(\lambda)$ 
do not depend on parameter $\lambda$, 
and they are close to the theoretical value: 
$\theta_2=0.71$ with respect to errors, as shown 
in Fig. \ref{fig:exponent8T}.

%
%
\subsection{Distribution functions $f(x; i, j; N)$ of the distance 
between vertices $i$ and $j$ of SAW and exponent $\theta(i, j)$}

\begin{figure}[htpb]
\begin{center}
\includegraphics[width=9cm,clip]{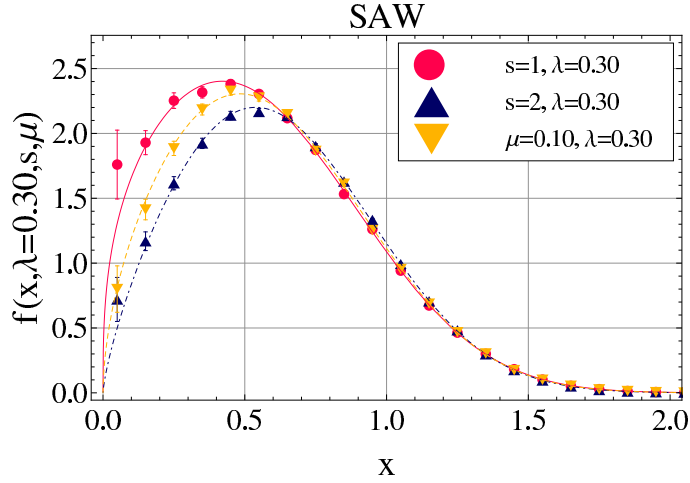}
\caption{Distribution functions $f_s(x; \lambda, N)$ with $\lambda=0.30$ 
for $s=1,2$, 
and $f(x; i,j; N)$ with $i=\mu \lambda N$ and $j=(1+\mu)\lambda N$ 
where $\lambda=0.30$ and $\mu=0.10$,  
are plotted against normalized distance $x$ between two vertices.  
Here $N=8,000$. Each curve has 20 data points with $0.05 \le x \le 1.95$. }
\label{fig:lam-mu}
\end{center}
\end{figure}

In Fig. \ref{fig:lam-mu} distribution functions 
$f_s(x; \lambda, N)$ with $\lambda=0.30$ and $N=8, 000$ 
for $s=1, 2$  and distribution function  
$f(x; i,j; N)$ with $i=\mu \lambda N$ and $j=(1+\mu)\lambda N$  
with $\lambda=0.30$ and  are plotted against 
normalized distance $x$. Here we recall $N=8, 000$. 
In the small $x$ region, the fitting curve of $\mu=0.1$ 
is located between those of $s=1$ and $s=2$. 
In the large $x$ region,  the three fitting curves 
overlap each other for $s=1, 2$ and $\lambda=0.3$. 
We therefore suggest that the asymptotic behavior for large $x$ 
is the same among the three cases of $f_s(x; \lambda, N)$ with $\lambda=0.30$  
for $s=1, 2$ and $f(x; i,j ; N)$ with $i=\mu \lambda N$ and $j=(1+\mu)\lambda N$ with $\lambda=0.30$.

In Figs. \ref{fig:exponent8T} we observe 
that as parameter $\lambda$ increases up to $\lambda=1$ 
the exponents $\theta_s(\lambda)$ for $s=1, 2$ are decreasing and become 
close to the value of exponent $\theta_0$.   
Here we remark that at $\lambda=1$, the distance between the 
two vertices is nothing but the end-to-end distance. 
Therefore, we may expect that the values of exponents $\theta_s(\lambda)$ 
for $s=1, 2$ approach the value of $\theta_0$ when we send $\lambda$ to 1.

Let us now introduce exponent $\theta(i,j)$ in order to 
describe the short-distance correlation of  
distribution function $f(x; i, j;  N)$ 
for normalized distance $x$ between 
the $i$th and $j$ th vertices of an $N$-step SAW. 
We introduce the fitting formula 
for the distribution functions $f(x; i, j; N)$ as follows.  
\be 
f(x; i, j; N) = C_{i,j} \, x^{\theta(i, j)} 
\exp \left( - 
(D_{i,j} \, x)^{\delta} \right) \,  \label{eq:universal-f(ij)}
\ee
where $\delta= 1/(1-\nu)$. The constants $D_{i,j}$ and $C_{i,j}$ are given by 
\bea 
D_{i,j} & = & \sqrt{\frac {\Gamma((5+ \theta(i,j))/\delta) }
                       {\Gamma((3+ \theta(i,j))/\delta) } } \, ,  \nonumber \\ 
C_{i,j} &  = &  {\frac {\delta} {\Gamma((3+\theta(i,j))/\delta)}} 
\left(  {\frac {\Gamma((5+ \theta(i,j))/\delta) }
                       {\Gamma((3+ \theta(i,j))/\delta) }} 
  \right)^{(3+\theta(i,j))/\delta} \, . \label{eq:CD(ij)} 
\eea
Here we recall that $x$ denotes the normalized distance: 
$x = r/R_{N}(i,j)$, where $R_{N}(i,j) = \sqrt{A_{i,j}} \, 
(\lambda N)^{\nu}$.

Let us express vertices $i$ and $j$ of a SAW 
in terms of parameters $\lambda$ and $\mu$ as   
\be 
i=\mu \lambda N \,,  \quad j=(1+ \mu) \lambda N. 
\ee
If we fix parameter $\mu$, vertices $i$ and $j$ satisfy the 
following relation  
\be 
j = i \left( 1+ \mu^{-1} \right) \, . \label{eq:j/i}
\ee
Here we recall that for a pair of vertices 
$i$ and $j$ of a SAW, the numbers 
$N_1$ and $N_2$ are expressed in terms of parameters $\lambda$ and $\mu$
by $N_1= \mu \, \lambda N$ and $N_2= \lambda N$, as shown in Fig. \ref{fig:SAW}.

\begin{figure}[htpb] 
\begin{center}
\includegraphics[width=10cm,clip]{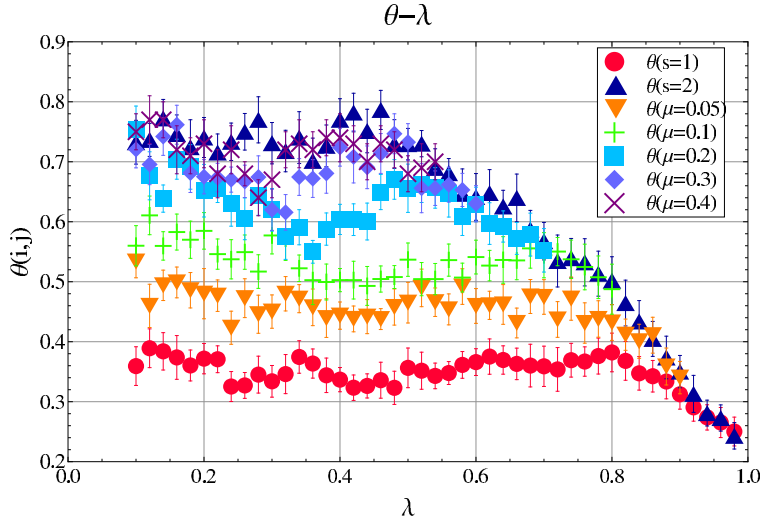}
\caption{Exponents $\theta(i, j)$ for the distance between 
the two points separated by $\lambda N$ steps where one point is located at 
$\mu \lambda N$ steps from an end point of SAW, i.e. $N_1=\mu \lambda N$. Here,  $N=8,000$.} 
\label{fig:exponent-middle}
\end{center} 
\end{figure}

We now show numerically that for a given value of $\mu$ 
exponent $\theta(i, j)$ is constant with respect to parameter 
$\lambda$ on the straight line segment: $0 \le \lambda \le 1/(1+2 \mu)$. 
In Fig. \ref{fig:exponent-middle} we observe that 
for given values of parameter $\mu$ such as 0.05, 0.1, 0.2, 0.3, 0.4,  
estimates of exponents $\theta(i, j)$ are independent of 
parameter $\lambda$ for $0.1 < \lambda < 0.5$.

Furthermore, in Fig. \ref{fig:exponent-middle} 
we observe crossover phenomenon such 
that if one of the two points separated by 
$\lambda N$ steps along SAW is close to an end point of SAW with less than 
$0.4 \lambda N$ (or $0.5 \lambda N$) steps 
(i.e. $\mu < 0.4$ or $\mu < 0.5$), 
then the value of exponent $\theta(i, j)$ changes 
from $\theta_2$ to $\theta_1$ as parameter $\mu$ approaches 0.

%
%
\subsection{Contour plot of exponents $\theta(i,j)$}

\begin{figure}
\begin{center}
\includegraphics[width=9cm,clip]{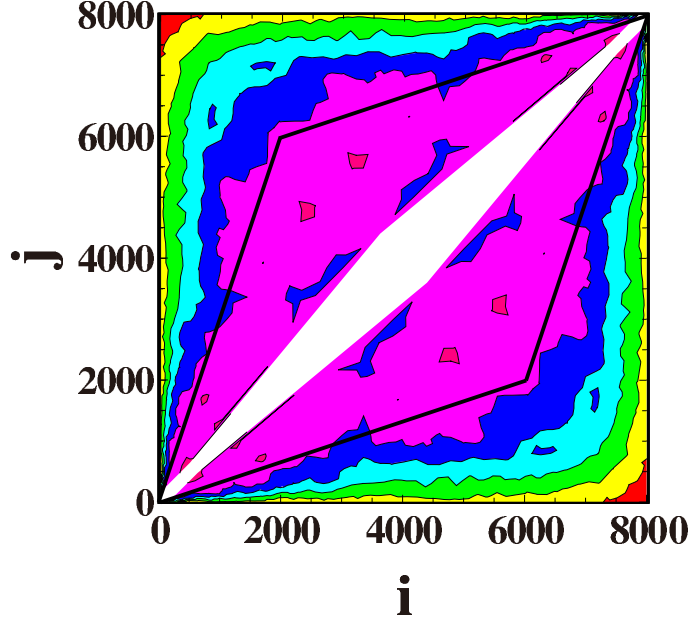}
\caption{Exponents $\theta(i,j)$  
for $0 < i, j < 8, 000$.  
In colored areas with red, yellow, green, cyan, blue, mazenta, and rose pink, 
we have $0.1 \times k \le \theta(i,j) < 0.1 \times (k+1)$ 
for $k=2, 3, \ldots, 8$ 
, respectively. 
The blank (or empty) region around the diagonal area 
where $|i-j|$ are small 
has no data due to poor statistics.  
} 
\label{fig:exponent}
\end{center} 
\end{figure}

Let us now show that exponent 
$\theta(i,j)$ as a function of $i$ and $j$ has a simple structure.  
The contours of exponent $\theta(i,j)$ 
are shown in Fig. \ref{fig:exponent}.  
The region where exponents $\theta(i,j)$ 
are larger than 0.7 and less than 0.8 is colored by mazenta. 
The mazenta region of $\theta(i,j)$ satisfying 
$0.7 \le \theta(i,j) < 0.8$ is approximately given 
by a rhombus with four black edges in Fig. \ref{fig:exponent}. 
We thus observe that the contour plot of $\theta(i,j)$ 
has a plateau region of $\theta(i,j) \approx \theta_2$ 
on the rohmbus in Fig. \ref{fig:exponent}.

The contour plot of  Fig. \ref{fig:exponent} clearly illustrate 
the following two observations:   
(i) In Fig. \ref{fig:exponent-middle} we observe that  
exponent $\theta(i, j)$ does not depend 
on parameter $\lambda$ in the region $0 \le \lambda \le 1/(1+2 \mu)$;   
(ii) In Fig. \ref{fig:exponent-middle} 
we observe crossover phenomenon: from $\theta_2$ to $\theta_1$ 
as $\mu$ approaches 0,  
and then from $\theta_1$ to $\theta_0$ as $\lambda$ approaches 1.0.

The graph of $(i,j)$ for paramter $\lambda$ 
satisfying $0 \le \lambda \le 1/(1+2 \mu)$ 
is given by the straight line with gradient  $1+ \mu^{-1}$ from the 
origin (0,0) to the crossing point $(i_c, j_c)$ in the graph of $i+j=N$. 
The crossing point $(i_c, j_c)$ is given by 
\be 
(i_c,j_c)=({\frac {\mu N} {1 +2 \mu}}, {\frac {(1+\mu)N} {1+2\mu}} ) .     
\ee
In Fig. \ref{fig:exponent} we observe 
that the value of exponent $\theta(i,j)$ is constant
on the line segment from the origin to the crossing point $(i_c,j_c)$. 
It depends on parameter $\mu$. Here we recall eq. (\ref{eq:j/i}).

For instance, the line of $\mu=0.5$ corresponds to the straight line from 
the origin (0,0) to (2000, 6000) in the coordinate of $(i,j)$ with 
$N=8000$. It is given by one of the four edges of the mazenta 
rohmbus where we have $\theta(i,j) \approx \theta_2$ 
in Fig. \ref{fig:exponent}. If $\mu$ becomes smaller than 0.5, then 
the gradient of the straight line increases and  $\theta(i,j)$ 
becomes smaller than $\theta_2$. Finally, we have $\theta(i,j)=\theta_1$ 
at $\mu=0$. Around at the two edges of (0, 8000) and (8000, 0) 
we have $\theta(i,j)=\theta_0$.

We now give an approximate expression for 
exponent $\theta(i,j)$ as a function of $i$ and $j$ as follows.    
For $0 \le \lambda \le 1/(1+2\mu)$  we have 
\be 
\theta(i,j) = 
\left\{ 
\begin{array}{cc}   
\theta_0^{(MC)} & \mbox{for} \, \mu=0 \, \mbox{and} 
\, 0.8 < \lambda \le 1.0 \\
\theta_1^{(MC)} & \mbox{for} \, \mu=0 \, \mbox{and} 
\, 0 \le \lambda  \le 0.8 \\ 
\theta_1^{(MC)} + \displaystyle{\frac {4 \mu} {1 + 2\mu}} \, 
(\theta_2^{(MC)} - \theta_1^{(MC)})  & \mbox{for} \, \, 0 < \mu \le 0.5 \\
\theta_2^{(MC)}   &  \mbox{for} \, \, 0.5 \le \mu < \infty \, . 
\end{array}
\right. 
\ee
Here, estimates $\theta_s^{(MC)}$ for $s=0,1,2$ are 
given in eq. ({\ref{eq:theta-MC}).

%
%
\subsection{Correlation functions through exponents $\theta(i, j)$ of SAW}

The estimates of short-distance exponents $\theta(i,j)$ shown  
in Fig. \ref{fig:exponent} are useful for constructing 
various quantities of SAW. 
In fact, formula (\ref{eq:universal-f(ij)}) has only 
two parameters $\theta(i,j)$and $\delta$ (or $\nu$), 
and the probability distribution function $p(r; i, j; N)$ is determined 
if we give  parameter $A_{i,j}$ (or $R_{N}(i,j)$) 
in addition to $\theta(i,j)$ and $\delta$.   

For instance, the pair correlation function of SAW, 
$g(r)$, is given by 
the sum of the distribution functions of the distance between two vertices 
$i$ and $j$, $p(r; i, j; N)$, over all vertices $i$ and $j$ of SAW. 
\be 
g(r)  = {\frac 1 N}  \sum_{i=1}^{N} \sum_{j=1}^{N}  p(r; i, j; N) \, .   
\ee
In terms of parameters $\lambda$ and $\mu$ we have  
\be
g(r)  = 4N \int_{0}^{\infty} d \mu  \int_{0}^{1/(1+2\mu)} 
\lambda d \lambda \, p(r; i, j; N) \, .   
\ee

We define the static structure factor of SAW, $g({\bm q})$,  by 
\be 
g({\bm q})= {\frac 1 N} \sum_{i=1}^{N} \sum_{j=1}^{N} 
\langle \exp \left(i {\bm q} \cdot ({\bm r}_i - {\bm r}_j) \right) \rangle .  
\ee
Assuming the rotational symmetry we have 
\be 
\langle \exp \left(i {\bm q} \cdot ({\bm r}_i - {\bm r}_j) \right) \rangle  
= \int_0^{\infty} {\frac {\sin qr} {qr}}  \, \, 
p(r; i,j; N) \, 4 \pi r^2 dr \, , 
\ee
where $r=|{\bm r}_i - {\bm r}_j|$ and  $q=|{\bm q}|$. 
We  have 
the following expresion of the static structure factor of SAW:  
\bea 
g(q) & = & {\frac 1 N} \int_{0}^{N} dm \int_{0}^{N} dn 
\int_{0}^{\infty} {\frac {\sin qr} {qr}} \, p(r; m, n; N)  
4 \pi r^2 dr  \nonumber \\ 
& = & 4N \int_{0}^{\infty} d\mu  
\int_{0}^{1/(1+2\mu)} \lambda d \lambda 
\int_{0}^{\infty} {\frac {\sin qr} {qr}} \, p(r; \lambda, \mu ; N)  
4 \pi r^2 dr \, .  
\eea

Furthermore, the diffusion coefficient of a linear polymer in a good solvent, 
$D_{G,L}$,  can be evaluated through 
the probability distribution functions of the distance between two points 
through the method of Kirkwood's approximation \cite{Doi-Edwards}.  
In the method $D_{G,L}$ is given by taking the sum of the ensemble average of inverse distance between two points of SAP over all pairs.    
\be 
D_{G,L}  =   {\frac {k_B T} {6 \pi \eta_s N^2}}  
\sum_{i=1}^{N} \sum_{j=1}^{N} \,  
\langle {\frac 1 {|{\bm r}_i - {\bm r}_j|} } \rangle \, . 
\ee
Here $\eta_s$ denotes the solvent viscosity.  
We then evaluate the ensemble average of the inverse distance in terms of 
the probability distribution function, as follows. 
\bea 
D_{G,L} & = & 
 {\frac {k_B T} {6 \pi \eta_s N^2}}  
\sum_{i=1}^{N} \sum_{j=1}^{N} \,  
\int_{0}^{\infty}  {\frac 1 {r} } \,  
p(r; i, j; N) \, 4 \pi r^2 dr \nonumber \\   
& = & 4N \int_{0}^{\infty} d \mu  \int_{0}^{1/(1+2\mu)} 
\lambda d \lambda \, 
\int_{0}^{\infty}  \,  
{\frac 1 r} \,  p(r; i, j; N) \, 4 \pi r^2 dr \, . 
\eea

%
%
\section{Scaling behavior of interchain correlation of SAP}

%
%
\subsection{Mean-square radius of gyration for SAP of cylindrical segments under a topological constarint}

Let us now show the data of the mean-square radius of gyration for cylindrical 
SAP under a topological constraint of type $K$. 
We denote it by $R_{g, K}^2$, briefly.  
Here we recall that  SAP consists of $N$ cylindrical segments 
with radius $r_{ex}$ of unit length.  
In Fig. \ref{fig:SAP_RgAll} 
the mean-square radius of gyration of cylindrical SAP 
under no topological constraint, $R_{g, All}^2$,  
is plotted for several different values of radius $r_{ex}$ 
with several numbers $N$ of segments upto $N=3,000$. 
The theoretical curves given by 
the formula $R_g^2 = A N^{2 \nu} (1 + B/N )$ are 
shown in Fig. \ref{fig:SAP_RgAll} together with the data points 
obtained by simulation. 
The $\chi^2$ values per datum are shown in Table \ref{tab:fit-SAP}. 
The $\chi^2$ values are small in the cases of $r_{ex}=0.0$ (i.e., the ideal case) and $r_{ex}=0.10$. Here, in the latter case the excluded volume 
has the largest value.

In the case of $r_{ex}=0.10$,  the estimate of exponent $\nu$ 
is numerically close to the exponent of SAW, 
as shown in Table \ref{tab:fit-SAP}.  
We suggest that only for the case of $r_{ex}=0.10$,  
 the SAP is long enough so that the excluded volume is fully effective. 
We shall also confirm it through 
fitting curves to the distribution functions of the distance 
between two points in subsection 4.2.

\begin{figure}[htpb] 
\begin{center}
\includegraphics[width=9cm,clip]{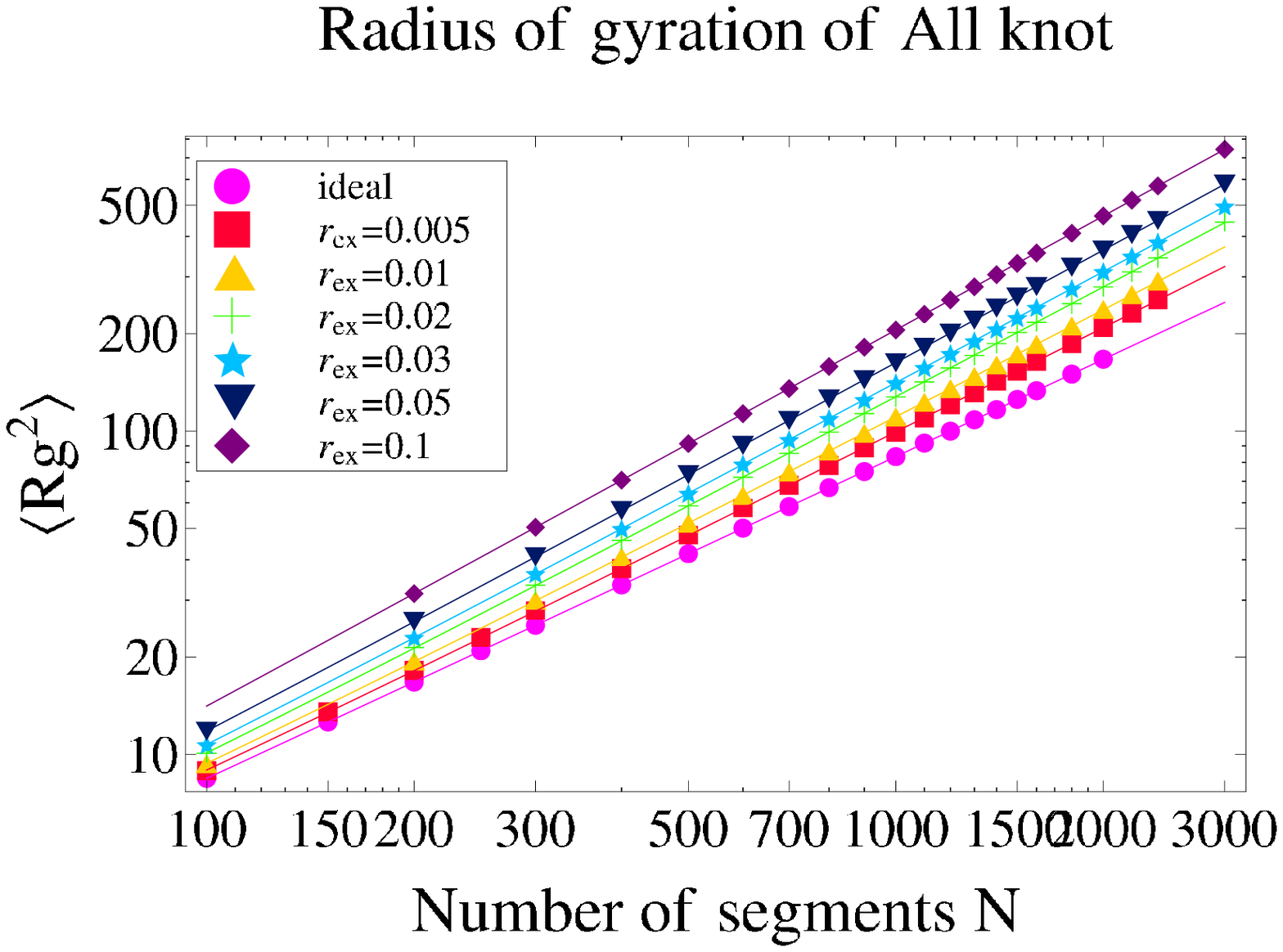}
\caption{Double-logarithmic plot of the mean-square radius of gyration  
of SAP consisting of $N$ cylindrical segments with radius $r_{ex}$ 
of unit length under no topological constraint 
$\langle R_{g}^2 \rangle_{All}$. 
For the cases of $r_{ex}$= 0.0, 0.05, 0.01, 0.02, 0.03, 0.05 and 0.1 are 
plotted with purple filled circles, red filled squares, 
yellow filled triangles, green crosses, light blue stars, dark blue triangles, 
and purple filled diamonds, respectively.  
} 
\label{fig:SAP_RgAll}
\end{center} 
\end{figure}

\begin{table}[htbp] \begin{center}
\begin{tabular}{c|cc|cc|cc|c} 
$r_{ex}$ & $A$ & & $\nu$ & & $B$ & 
& $\chi^2$/datum \\ \hline 
0 & 0.0833 & $\pm$ 0.0004 & 0.5000 & $\pm$ 0.0003 & 1.2 & $\pm 0.2$ & 1.04  
\\
0.005 & 0.0584 & $\pm$ 0.0006 & 0.5379 & $\pm$ 0.0008 & 8.0 & $\pm 0.6$ & 7.93 
\\
0.01 & 0.0529 & $\pm$ 0.0006 & 0.5528 & $\pm$ 0.0007 & 9.0 & $\pm 0.6$ & 7.59 
\\
0.02 & 0.0504 & $\pm$ 0.0005 & 0.5667 & $\pm$ 0.0006 & 8.5 & $\pm 0.5$ & 7.99 
\\
0.05 & 0.0538 & $\pm$ 0.0003 & 0.5798 & $\pm$ 0.0003 & 5.3 & $\pm 0.3$ & 2.57 
\\
0.1 & 0.0623 & $\pm$ 0.0003 & 0.5862 & $\pm$ 0.0004 & 2.7 & $\pm 0.5$ & 0.421 
\\
\hline 
\end{tabular}
\caption{Best estimates of the fitting formula: 
$R_{g}^2 = A N^{2 \nu} (1 + B/N)$ for the mean-square radius of gyration 
for SAP consisting of $N$ cyllindrical segments with raidus $r_{ex}$ of unit length under no topological constraint $\langle R_g^2 \rangle_{All}$ .  } \label{tab:fit-SAP}
\end{center}
\end{table} 

In Fig. \ref{fig:SAP_RgTrivial}
the ratio of the mean-square radius of gyration 
of the cylindrical SAP with the trivial knot ($0_1$), $R^2_{g, 0_1}$, 
to that of under no topological constraint ($All$), $R^2_{g, All}$, 
is plotted against the number of nodes $N$ 
for several different values of cylindrical radius $r_{ex}$. 
The ratio is always larger than 1.0 
except for the case of $r_{ex}=0.1$. 
Furthermore, the ratio decreases as the cylindrical radius increases. 
Here we remark that the values of $R^2_{g, All}$ for the different values of 
cylinder radius $r_{ex}$ are given in Fig. \ref{fig:SAP_RgAll}.

The ratio of the mean-square radius of gyration 
of the cylindrical SAP with the trefoil knot ($3_1$), $R^2_{g, 3_1}$,  
to that of under notopological constraint, $R^2_{g, All}$,  
is plotted in Fig. \ref{fig:SAP_Rg31} 
against the number of nodes $N$ for different values of 
cylindrical radius $r_{ex}$. 
The ratio is larger than 1.0 
for the cases of small values of $r_{ex}$ and 
large $N$.  It decreases as the cylindrical radius increases.

\begin{figure}[htpb] 
\begin{center}
\includegraphics[width=8cm,clip]{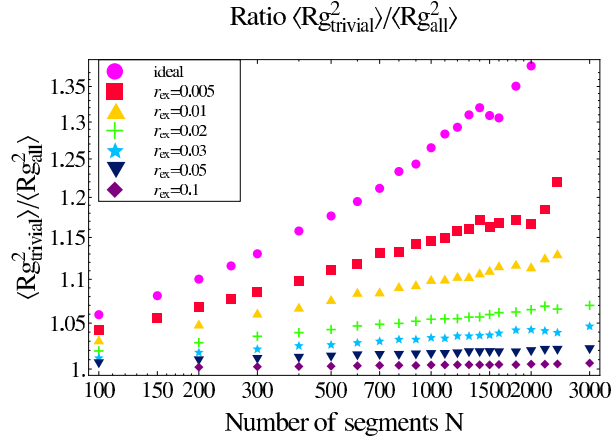}
\caption{Double-logarithmic plot of the ratio 
of the mean-square radius of gyration  
for the $N$-noded SAP with trivial knot ($0_1$) to that of no topological constraint (including all knots) against the number of nodes $N$ for 
various different values of cylindrical radius $r_{ex}$. 
Error bars are not shown in the figure. 
} 
\label{fig:SAP_RgTrivial}
\end{center} 
\end{figure}

We thus observe {\it topological swelling} 
in Figs. \ref{fig:SAP_RgTrivial} and \ref{fig:SAP_Rg31} \cite{SAP02}. 
In the cases when the cylindrical radius $r_{ex}$ is small, 
the mean-square radius of gyration of cylindrical SAP with 
a fixed knot type becomes larger than that of no topological constraint 
for large enough $N$; i.e., the ratio 
$R^2_{g, K}/R^2_{g, All}$ becomes larger than 1.0 if $N$ is large enough.
Here $K$ denotes a knot type.   
We consider that topological swelling occurs  
since entropic repulsive forces appear effectively among segments of the SAP 
under a topological constraint of a fixed knot \cite{desCloizeaux-Let,SAP02}. 
Here we remark that the ratios of the mean-square radii of gyration 
 of $N$-noded cylindrical SAP under a topological cnstraint, 
$R^2_{g, K}/R^2_{g, All}$,  were evaluated for the trivial 
and trefoil knots in Ref. \cite{SAP02}, 
although the number of $N$ was limited upto $N = 1,000$.

We also observe in Fig. \ref{fig:SAP_Rg31}  
that ratio $R^2_{g, 3_1}/R^2_{g, All}$ is smaller than 1.0 
for any value of radius $r_{ex}$ if $N$ is smaller than 200. 
It is due to the finite-size effect: The polymer chain is short so that  
the size of the polymer making the trefoil knot is rather small. 
However, if the chain is long enough, the necessary number of segments to make 
the trefoil knot becomes much smaller than the total number of segments $N$,  
and the rest of the chain becomes as large as SAP of the trivial knot. 
Therefore,   the ratio 
$\langle R^2_{g, 3_1} \rangle/\langle R^2_{g, All} \rangle$,  
becomes larger than 1.0 for large $N$.

\begin{figure}[htpb] 
\begin{center}
\includegraphics[width=8cm,clip]{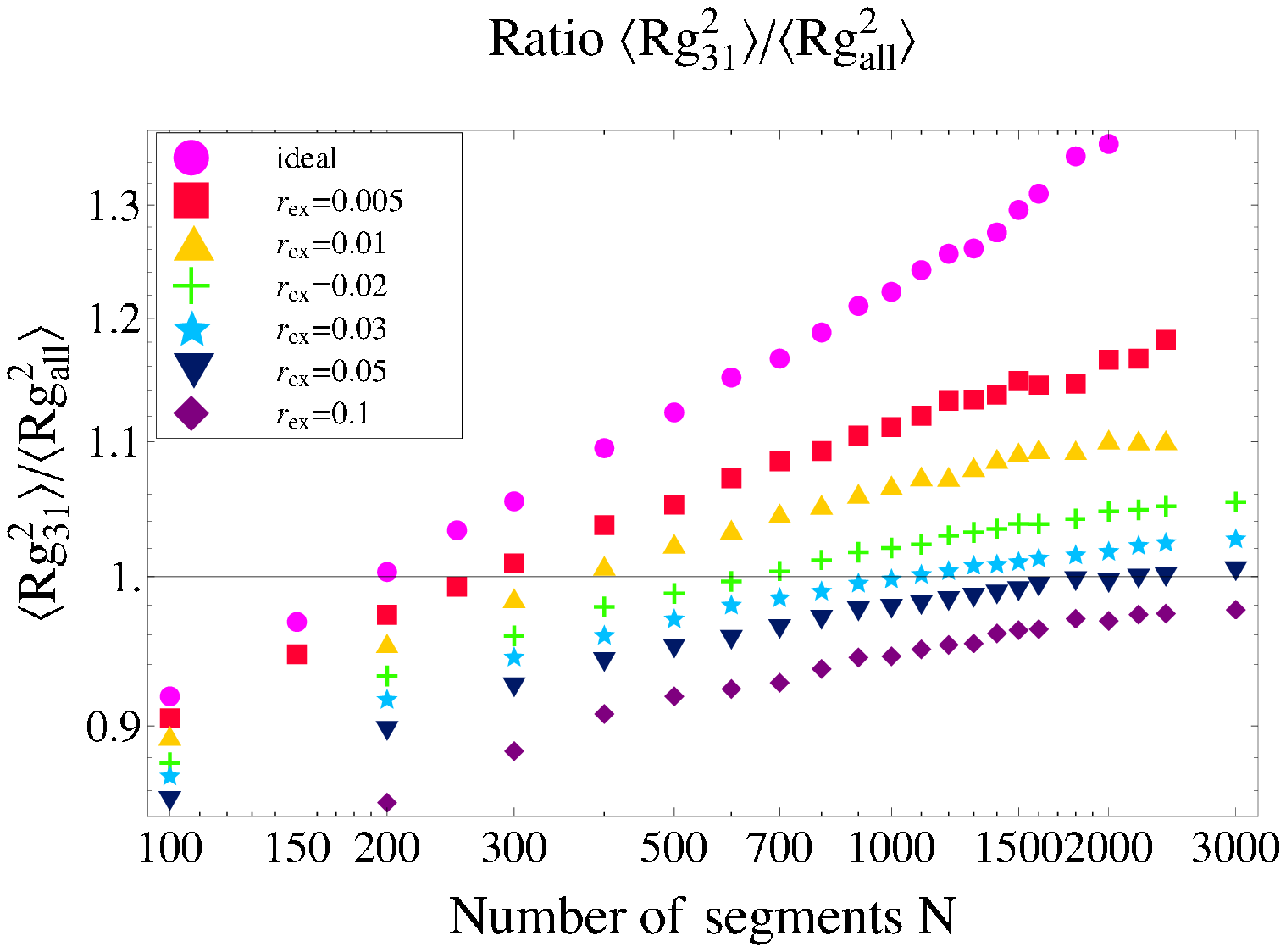}
\caption{Double-logarithmic plot of the ratio 
of the mean-square radius of gyration  
for the $N$-noded SAP with the treofil knot ($3_1$) to 
that of no topological constraint (including all knots) 
against the number of nodes $N$ for various different values 
of cylindrical radius $r_{ex}$. 
Error bars are not shown in the figure.  
} 
\label{fig:SAP_Rg31}
\end{center} 
\end{figure}

Topological swelling occurs only if the excluded volume 
is small \cite{Swelling}. 
For $r_{ex}=0.05$, the ratio $\langle R^2_{g, 3_1} \rangle/\langle R^2_{g, All} \rangle$,  becomes larger than 1.0 only at $N=3,000$, 
as shown in  Fig. \ref{fig:SAP_Rg31}. 
We may therefore consider that the excluded volume effect is not compatible 
with the topological entropic repulsions among segments of SAP 
under a topological constraint.

%
%
%
\subsection{Probability distribution functions 
of the distance between two nodes of SAP}

Let us consider the probability disrtibution function of the distance 
between two nodes $i$ and $j$ of SAP with knot type $K$ 
consisting of $N$ cylindrical segments 
with radius $r_{ex}$ and of unit length. 
We define it by 
\be 
p_K({\bm r}; i, j; N) =  
\langle \delta({\bm r}- ({\bm R}_j - {\bm R}_i))  \rangle_K \, . 
\ee
Here the symbol $\langle \cdot \rangle$ denotes 
the average over all possible configurations of cylindrical SAP 
of $N$ nodes havine knot type $K$.  For the case of no topological constraint, 
we denote $K$ as {\it All}, which suggests that all knots are included.

Due to the cyclic symmetry, the distribution function 
$p_K({\bm r}; i, j; N)$ depends 
only on the distance $|i-j|$. 
Let us introduce parameter $\lambda$ for SAP by 
\be 
\lambda = |j-i|/N 
\ee
We thus express the probability disrtibution function of the distance 
between two points $i$ and $j$ of SAP, 
as  $p_K(r; \lambda, N)$. 

Let us denote by $R_{N,K}(\lambda)$ 
the square root of the mean square distance 
between $i$ and $j$ of $N$-noded SAP with knot type $K$ where $i$ and $j$ are
separated by $\lambda N$ steps. It is given by the following:   
\be 
R_{N, K}(\lambda) = \sqrt{\langle r_{i, j}^2 \rangle_K } \, . 
\ee
For the distance between the two nodes $i$ and $j$, $r_{i,j}$,  
we introduce the normalized distance $x_{i,j}$ by 
\be 
x_{i,j}= r_{i,j}/R_{N,K}(\lambda) \, .  
\ee
We denote by $f_K(x; \lambda, N)$ 
the distribution function of normalized distance $x$ 
between two nodes $i$ and $j$ of $N$-noded SAP with knot $K$ 
consisting of  cylindrical segments with radius $r_{ex}$, 
where $i$ and $j$ are separated by $\lambda N$ steps. 
It is expressed in terms of the probability distribution function   
$p_K(r; \lambda, N)$ as follows. 
\be 
f_K(x; \lambda, N) 
= 4 \pi R^3_{N,K}(\lambda) \, p_K(x R_{N,K}; \lambda, N) \, . 
\ee
Here we recall $|i-j| = \lambda N$.

\begin{figure}[htpb] 
\begin{center}
\includegraphics[width=6cm,clip]{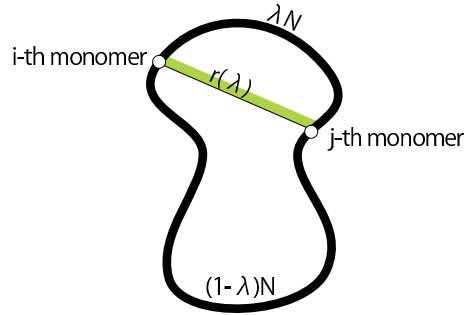}
\caption{Distance $r(\lambda)$ between the $i$th and $j$th vertices 
of a long Self-Avoiding Polgon (SAP) of $N$ steps, 
where $1 \le i < j \le N$ and $N \gg1$.  In terms of parameter $\lambda$ 
we express the difference $|i-j|$ as $|i-j|= \lambda N$ 
for $0 \le \lambda \le 1/2$. 
} 
\label{fig:SAP}
\end{center} 
\end{figure}

Let us introduce the formula for fitting curves to the data of   
the probability distribution function of the normalized distance between 
two segments separated by $\lambda N$ steps, 
$f(x; \lambda, N)$, as follows.  
\be 
f_K(x; \lambda, N) = C_K(\lambda) \, x^{\theta_K(\lambda)} 
\exp \left( - 
(D_K \, x)^{\delta} \right) \,  \label{eq:universal-f-SAP}
\ee
where $\delta= 1/(1-\nu)$. The constants $D$ and $C$ are given by 
\bea 
D_K & = & \sqrt{\frac {\Gamma((5+ \theta_K)/\delta_K) }
                       {\Gamma((3+ \theta_K)/\delta_K) } } \, ,  \nonumber \\ 
C_K &  = &  {\frac {\delta} {\Gamma((3+\theta_K)/\delta_K)}} 
\left(  {\frac {\Gamma((5+ \theta_K)/\delta_K) }
                       {\Gamma((3+ \theta_K)/\delta_K) }} 
  \right)^{(3+\theta_K)/\delta_K} \, . \label{eq:CD-SAP} 
\eea
Here we recall that $x$ denotes the normalized distance: 
$x = r/R_{N, K}(\lambda)$, where $R_{N, K}(\lambda) = \sqrt{A_K} \, 
(\lambda N)^{\nu_K}$.

\begin{figure}[htpb] 
\begin{center}
\includegraphics[width=8cm,clip]{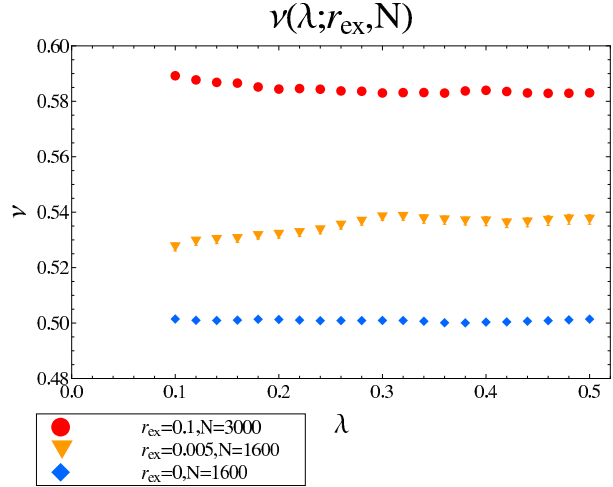}
\caption{Estimates of exponent $\nu_{All}(\lambda)$ evaluated 
by applying formula (\ref{eq:universal-f-SAP}) to 
the distribution function of the distance between two vertices  
of  SAP with $N$ cylindrical segments of radius $r_{ex}$ 
under no topological constraint. } 
\label{fig:nu_50sampling}
\end{center} 
\end{figure}

We have evaluated exponent $\nu_K(\lambda)$ of the $N$-noded cylindrical 
SAP with topological condition $K$ 
of radius $r_{ex}$ for various values of $r_{ex}$ such as 
$r_{ex}$=0, 0.005, 0.01, 0.02, 0.05, and 0.1, and various numbers of $N$ such as$N$=400, 800, 1,600, 2,000, 3,000.  
We have applied  formula (\ref{eq:universal-f-SAP}) to the data of 
the distribution function of the normalized distance 
between two nodes of the cylindrical SAP,    
and obtained the best estimates of exponents $\delta_K(\lambda)$ 
for each value of $\lambda$ from the fitting curves to the data. 
We then calculated $\nu_K(\lambda)$  from the estimates of  $\delta_K(\lambda)$
Here, we have considered the three topoogical conditions, 
the trivial knot ($0_1$), the trefoil knot ($3_1$), 
and no topological constraint (``$All$''), 
and for 25 values of $\lambda$ from 0.02 to 0.5 by 0.02.  
The $\chi^2$ values per datum are given by less than or equal to 1.0 or  2.0 
for all the fitting curves. Thus, we conclude that 
the fitting curves are good.

For an illustration, 
in Fig. \ref{fig:nu_50sampling}, the estimates of $\nu_{All}(\lambda)$ 
with no topological constraint (i.e. $K=All$) are plotted against $\lambda$ 
for the three cases: the thick case of $r_{ex}=0.10$ and $N=3,000$ (filled red circles), the thin case of $r_{ex}=0.005$ and $N=1,600$ (downward orange triangles) and the ideal case of $r_{ex}=0.0$ 
and $N=1,600$ (filled dark-blue diamonds). 
Here we plot the ideal case for reference. 

In the thick case ($r_{ex}=0.10$ and $N=3,000$), 
the value of  $\nu_{All}(\lambda)$ is almost consistent with the exponent of SAW, $\nu_{SAW}=0.588$ over all range of $\lambda$. It is 
However, in the thin case ($r_{ex}=0.005$ and $N=1,600$), 
the value of  $\nu_{All}(\lambda)$  is rather smaller 
than the exponent of SAW, $\nu_{SAW}$. Moreover, 
in the ideal case ($r_{ex}=0.0$ and $N=1,600$)  
the estimates of  $\nu_{All}(\lambda)$ almost equal to 0.5 for all values of 
$\lambda$.

Thus, as far as the excluded volume effect of SAP is concerned, 
we conclude that for $r_{ex}=0.10$, SAW of $N=3,000$ is large enough to see the effect of excluded volume, while for $r_{ex}=0.005$ and $N=1,600$, 
SAW of $N=1600$ is not large enough to see it.

%
%
\subsection{Exponents $\theta_K(\lambda)$ of short-distance correlation of 
SAP with knot $K$}

%
%
\subsubsection{The case of thick cylindrical SAP}

\begin{figure}[htbp]
\begin{center}
\includegraphics[width=8cm,clip]{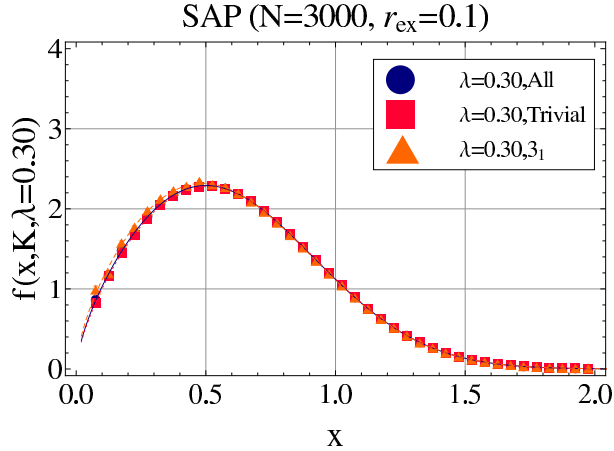}
\caption{Data points and fittung curves of the 
distribution functions $f_K(x; \lambda, N)$ of normalized distance $x$ 
between two nodes of cylindrical SAP of $N=3,000$ with radius 
$r_{ex}=0.1$ (thick cylinders) at $\lambda=0.3$ 
for three topological conditions: the trivial knot ($0_1$), 
the trefoil knot ($3_1$) and no toplogical constaint ($All$).
} \label{fig:fK010}
\end{center}
\end{figure}

We plotted in Fig. \ref{fig:fK010} the data points of distibution functions 
$f_K(x; \lambda, N)$ of normalized distance $x$ 
between two nodes of cylindrical SAP of $N=3,000$ consisting 
of cylindrical segments of radius 
$r_{ex}=0.1$ of unit length (thick cylinders). 
Here, the two nodes are separated by 
$\lambda N$ steps with $\lambda=0.3$. We consider   
three topological conditions: the trivial knot ($0_1$), 
the trefoil knot ($3_1$) and no toplogical constaint ($All$). 
We recall that the fitting curves are given 
by formula (\ref{eq:universal-f-SAP}) with 
two parameters $\theta_K$ and $\delta_K$.

\begin{table}[htbp] \begin{center}
\begin{tabular}{|c|rl|rl|rl|}   
\hline   
$K$ & No constraint  & ($All$) & trivial knot & ($0$) &  trefoil knot & ($3_1$)  \\ \hline 
$\theta_K$  &  0.664   & $\pm 0.004$   &  0.679    & $\pm 0.004$  & 0.623  & $\pm 0.009$  \\ \hline 
$\nu_K$ & 0.583 & $\pm 0.001$ &  0.583  &  $\pm 0.001$   & 0.581 
& $\pm 0.002$  \\ \hline 
$\chi^2$/datum & 1.19 & & 1.07  & & 1.24  & \\ \hline 
 \end{tabular} 
\caption{Estimates of fitting parameters and the $\chi^2$ value per datum 
for the fitting curves in Fig. \ref{fig:fK010}: 
$r_{ex}=0.1$, $\lambda=$ 0.3 and $N=3000$. }
 \label{tab:fK010}
\end{center}
\end{table}

For all the three topological conditions 
the $\chi^2$ values per datum are small.  
We thus find that the fitting curves to the distribution functions are good. 
The fitting parameters are listed in Table \ref{tab:fK010}. 
Here we recall that we have observed in Fig. \ref{fig:SAP_RgAll} 
the excluded volume effect appears clearly 
for SAP with $N=3,000$ segments in  
the case of $r_{ex}=0.1$, i.e., the thick cylindrical SAP.

\begin{figure}[htbp]
\begin{center}
\includegraphics[width=8cm,clip]{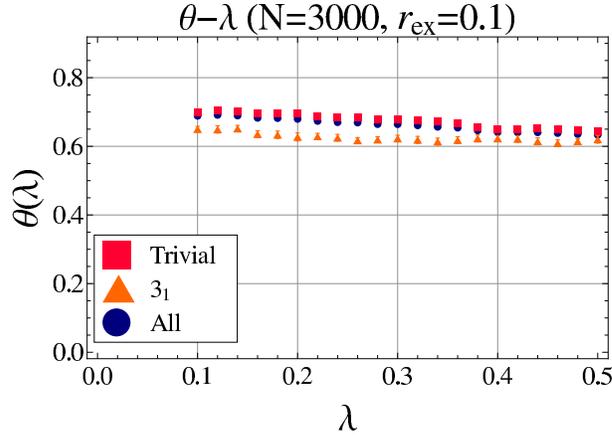}
\caption{Estimates of $\theta_K(\lambda)$ against parameter $\lambda$
for cylindrical SAP with radius $r_{ex}=0.1$ of $N=3000$. 
Here $K$ is given by the trivial knot ($0_1$), 
the trefoil knot ($3_1$) and no toplogical constaint ($All$).}
\end{center}
\end{figure}

For the thick cylindrical SAP  we plotted in Fig. \ref{fig:0100nu_K} 
the estimates of exponent $\theta_K(\lambda)$ 
against $\lambda$ with  
25 values of parameter $\lambda$ for three topological conditions: the trivial knot ($0_1$), the trefoil knot ($3_1$) and no toplogical constraint ($All$). 
All the estimates of exponent $\theta_K(\lambda)$ 
are roughly the same such as $\theta(\lambda) \approx 0.7$. 
 In each topological condition $K$ 
the estimate of  $\theta_K(\lambda)$ is almost constant with respect 
to parameter $\lambda$. 
Furthermore, they do not depend on topological conditions $K$. 
The exponent for the trivial knot and the trefoil knot is rather cose 
to each other,  while that of the trefoil knot is smaller than others.  
that of the : $\theta_{0_1} \approx \theta_{All} > \theta_{3_1}$.   

\begin{figure}[htbp]
\begin{center}
\includegraphics[width=8cm,clip]{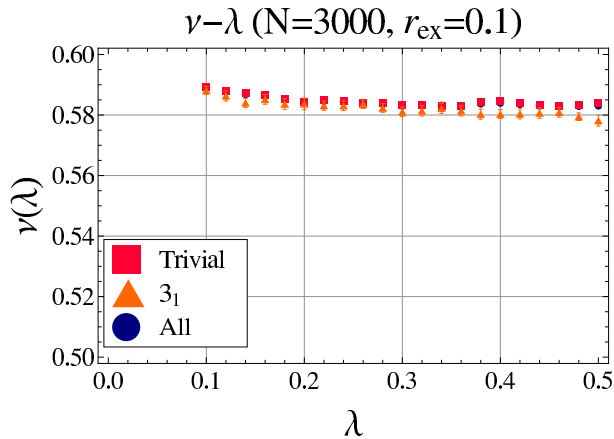}
\caption{Estimates of $\nu_K(\lambda)$ against parameter $\lambda$
for cylindrical SAP 
with radius $r_{ex}=0.1$ of $N=3000$. Here  
$K$ is given by the trivial knot ($0_1$), 
the trefoil knot ($3_1$) and no toplogical constaint ($All$).}
\label{fig:0100nu_K}
\end{center}
\end{figure}

For the thick cylindrical SAP  we also plotted in Fig. \ref{fig:0100nu_K} 
the estimates of exponent $\nu_K(\lambda)$  against $\lambda$ with  
25 values of parameter $\lambda$ for three topological conditions: the trivial knot ($0_1$), the trefoil knot ($3_1$) and no toplogical constaint ($All$).
All of them are numerically close to the value of exponent of SAW, 
$\nu_{\rm SAW}=0.588$. 
The estimates of exponent $\nu_K(\lambda)$ 
are independent of topological conditions. 
We observe that the exponent of $3_1$, $\nu_{3_1}$, 
is a little smaller than the other two cases.  
As a function of parameter $\lambda$, 
the estimate of exponent $\nu_K(\lambda)$ is almost constant 
for each of the three topological conditions $K$.

%
%
\subsubsection{The case of thin cylindrical SAP}

In the case of thin cylindrical SAP with radius $r_{ex}=0.005$  
we plotted in Fig. \ref{fig:fK0005} 
the data points of distibution functions 
$f_K(x; \lambda, N)$ of normalized distance $x$ 
between two nodes of cylindrical SAP of $N=1,600$ 
with parameter $\lambda=0.3$.  
Here we recall that the two nodes of SAP 
are separated by $\lambda N$ steps along the chain. 
We consider the three topological conditions: the trivial knot ($0_1$), 
the trefoil knot ($3_1$) and no toplogical constaint ($All$). 
The fitting curves given 
by formula (\ref{eq:universal-f-SAP}) are also plotted in Fig. \ref{fig:fK0005}, which are determined with two fitting parameters $\theta_K$ and $\delta_K$.

\begin{figure}[htbp]
\begin{center}
\includegraphics[width=8cm,clip]{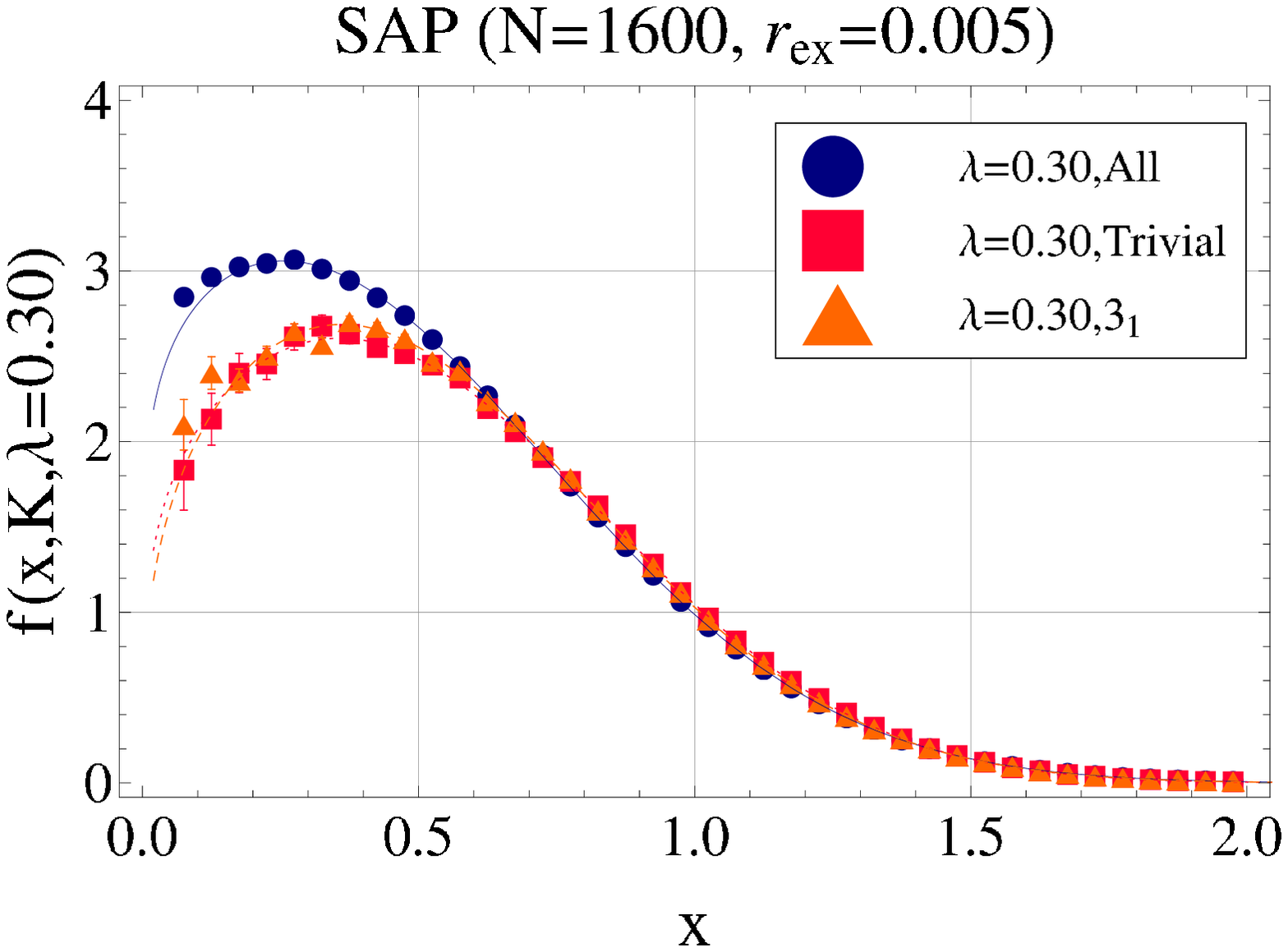}
\caption{Data points and fittung curves 
of the distibution functions $f_K(x; \lambda, N)$ of normalized distance $x$ 
between two nodes of cylindrical SAP of $N=1600$ 
with radius $r=0.005$ (thin cylinders) for topological conditions, 
such as the trivial knot ($0_1$), the trefoil knot ($3_1$) and 
no toplogical constaint ($All$) (thin cylinders). }
\label{fig:fK0005} 
\end{center}
\end{figure}

Also for the thin cylidrical SAP 
the $\chi^2$ values per datum are small 
for all the three topological conditions.  
We find that the fitting curves to the distribution functions are good. 
The fitting parameters are listed in Table \ref{tab:fK0005}. 

 The estimates of exponent $\theta_K$ in the thin case 
are much smaller than the case of thick SAP. For instance, 
we have $\theta _{all}=0.2$.  
It is maybe due to the fact that 
the excluded volume effect is not strong, yet.    
Here we recall that we have observed in Fig. \ref{fig:SAP_RgAll} 
the excluded volume effect does not clearly appear 
for SAP of $N=1600$ segments in the case of $r_{ex}=0.005$.

\begin{table}[htbp] \begin{center}
\begin{tabular}{|c|rl|rl|rl|}   
\hline   
$K$ & No constraint  & ($All$) & trivial knot & ($0$) &  trefoil knot & ($3_1$)  \\ \hline 
$\theta_K$  &  0.161   & $\pm 0.005$   &  0.26    & $\pm 0.02$  & 0.33  & $\pm 0.02$  \\ \hline 
$\nu_K$ & 0.538 & $\pm 0.001$ &  0.582  &  $\pm 0.004$   & 0.551 
& $\pm 0.004$  \\ \hline 
$\chi^2$/datum & 3.32 & & 0.492  & & 1.26  & \\ \hline 
 \end{tabular} 
\caption{Estimates of fitting parameters and the $\chi^2$ value per datum 
for the fitting curves in Fig. \ref{fig:fK0005}: 
$r_{ex}=0.005$, $\lambda=$ 0.3 and $N=1600$. }
 \label{tab:fK0005}
\end{center}
\end{table}

%
%
\subsection{Correlation functions through exponents $\theta(\lambda)$ }

We now discuss that the distribution function of 
the distance between two points of SAP with knot type $K$, 
$p_K(r; i, j; N)$,   
is useful for constructing various important quantities 
of knotted ring polymers in solution. 
Let us assume that the two points are separated 
by $\lambda N$ steps along the chain of the SAP. 
Due to the translational symmetry among the vertices of SAP 
along the chain,  the expressions of the physical quantities 
are much simpler than those of SAW. 
Here we remark that the structrue factor of dilute ring polymers have been 
studied numerically \cite{Vicari}.

The pair correlation function of SAP with knot $K$ is given by  
\bea 
g_K(r) &  = & {\frac 1 N}  \sum_{i=1}^{N} \sum_{j=1}^{N}  p_K(r; i, j; N)  
\nonumber \\  
&  = &  \sum_{j=1}^{N}   p_K(r; 0, j ; N)  \, . 
\eea
In terms of parameter $\lambda$  
we express it as a single integral as follows.   
\be
g_K(r)  = N \int_{0}^{1} d \lambda \, p_K(r; \lambda ; N) \, .   
\ee

%
%
\section{Diffusion constants of knotted SAP }

We now evaluate the diffusion coefficient of $N$-noded cylindrical SAP 
with a knot type $K$ in solution by  Kirkwood's approximation.   
\be 
D_{G,K}  =   
{\frac {k_B T} {6 \pi \eta_s N^2}}  
\sum_{i=1}^{N} \sum_{j=1}^{N} 
\langle \frac 1 {|{\bm R}_i - {\bm R}_j|} \rangle_K \, . 
\ee
Here we recall that $\eta_s$ denotes the solvent viscosity and 
$\langle \cdot \rangle_K$ denotes the average over all configurations of 
SAP with a given knot type $K$.

The estimates of the diffusion coefficient $D_{G,K}$ of 
 cylindrical SAP with radius $r_{ex}=0.1$ 
are plotted against the number $N$ of nodes 
for no topological constraint denoted $0_1$ 
and the trivial knot denote $0_1$ 
in Fig. \ref{fig:DGthick} 
in double-logarithmic scales.  
Here, we evaluated the ensemble average 
$\langle \frac 1 {|{\bm R}_i - {\bm R}_j|} \rangle_K$ 
by taking the sum over all the configurations of SAP with knot type $K$. 

For the two topological conditions, $K=All$ and $K=0_1$, 
the diffusion coeffiicents coincide with each other. Furthermore, they are well fitted by a staright line in the double-logarithmic scale.  
For the thick cylinder case, the topology of the majority of SAPs 
is given by the trivial knot,  and hence $D_{G,All}$ and 
$D_{G,0_1}$  have the same value.

\begin{figure}
\begin{center}
\includegraphics[width=8cm,clip]{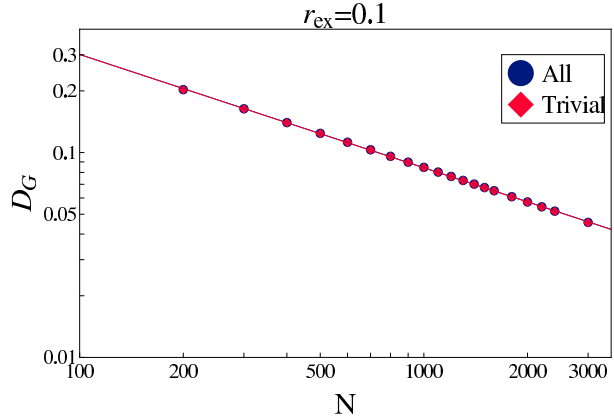}
\caption{Double logarithmic plot of diffusion coefficient of $N$ noded cylindrical SAP under topological condition $K$. The SAP 
consists of $N$ thick cylindrical segments with radius $r_{\rm ex}=0.1$ of unit length. Filled blue circle and filled red diamonds denote 
no topological constraint ($All$) and the trivial knot ($0_1$), 
respectively. The sum 
${\frac 1 {N^2}} \sum_{i=1}^{N} \sum_{j=1}^{N} \langle \frac 1 {|{\bm R}_i - {\bm R}_j|} \rangle_K$ is plotted. 
} 
\label{fig:DGthick}
\end{center} 
\end{figure}

\begin{figure}[htpb] \begin{center}
\includegraphics[width=9cm,clip]{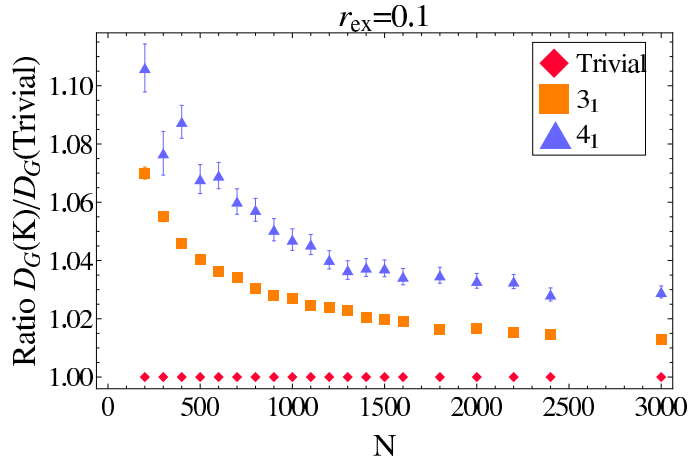}
\caption{Ratio of diffusion coefficients, $D_{G,K}/D_{G, 0_1}$, 
is plotted against the number of nodes $N$. 
Here $D_{G,K}$ denotes the diffusion coefficient of $N$-noded cylindrical SAP 
with $r_{\rm ex}=0.10$ for given topological condition $K$,   
and $0_1$ denotes the trivial knot.  } 
\label{fig:DG_Ratio(trivial)thick}
\end{center} \end{figure}

The ratio of diffusion coefficients $D_{G,K}/D_{G, 0_1}$ is plotted 
against the number of segments $N$ in Fig. \ref{fig:DG_Ratio(trivial)thick}.  
Here we recall that the diffusion coefficient of 
SAP under no topological constraint and that of the trivial knot coincides 
numerically very well. 

We observe in Fig. \ref{fig:DG_Ratio(trivial)thick} 
that the ratio gradually decreases with respect to the number of $N$.  
We suggest that it reaches an asymptotically constant value 
at some large value of $N$.  
Following renormalization group arguments 
we expect that the ratio should 
be universal and independent of details of models.  
It would be interesting to compare the ratio with experimental data in future.

The diffusion coeffiecient of the figure-eight knot 
has large error bars due to the number of SAP with the knot type is very small for the thick case with cylindrical radius $r_{ex}=0.1$.

In the method of Kirkwood's approximation we can express 
diffusion coefficient $D_{G,K}$ in terms of the probability distribution functions of the distance between two nodes of SAP. Here we take the sum over 
all pairs $i$ and $j$ of SAP as follows.  
\bea 
D_{G,K} & = &  
{\frac {k_B T} {6 \pi \eta_s N^2}}  
\sum_{i=1}^{N} \sum_{j=1}^{N} 
\int_{0}^{\infty} \, {\frac 1 r} \, p_K(r; i, j; N) \, 4 \pi r^2 . 
\eea
Considering the cyclic symmetry of SAP we reduce the double sum into the 
single sum, as follows.  
\be  
\sum_{i=1}^{N} \sum_{j=1}^{N} 
\int_{0}^{\infty} \, {\frac 1 r} \, p_K(r; i, j; N) \, 4 \pi r^2 
= N \sum_{j=1}^{N} \int_{0}^{\infty} \, {\frac 1 r} \, p_K(r; 0, j; N) 
\, 4 \pi r^2  \, . 
 \ee
In terms of $\lambda$ we have the following expression. 
\be 
D_{G,K} =  
{\frac {k_B T} {6 \pi \eta_s}} 
\int_{0}^{1} d \lambda \, \int_{0}^{\infty} \, 
{\frac 1 r} \,  p_K(r; \lambda; N) \, 4 \pi r^2 \, . 
\ee

%
%
\section{Concluding remarks}

We have shown that formula (\ref{eq:universal-f}) gives 
good fitting curves to the data of the probability distribution functions 
of the distance between two points of SAW, 
$f_0(x; N)$ and $f_s(x; \lambda, N)$ for $s=1, 2$,  
over a wide range of the normalized distance $x$ such as 
from $x=0.05$ to 1.95.

In the case of large $x$, the distribution functions  
have the same asymptotic behavior: 
$f_s(x; \lambda, N) \propto \exp(- x^{\delta})$  with $\delta=1/(1-\nu)$ 
for many different values of $\lambda$ such as from $\lambda$ = 0.10 to 0.98. 
Thus, exponent $\delta$ does not change for $s=0, 1, 2$ and for various values 
of $\lambda$. Moreover, exponent $\delta$ does not change for any vertices $i$ 
and $j$ of $N$-step SAW.

We evaluated the exponents $\theta_s$ 
which describe and characterize 
the short-distance behaviour of $f_s(x; \lambda, N)$, 
from the fitting curves to the data points from $x=0.05$ to 1.95, which is 
almost the entire region of $x$.  
The estimates of $\theta_1(\lambda)$  
are clealry smaller than the theoretical value $\theta_1^{(RG)}=0.46$ 
for $0.1 < \lambda < 0.8$; 
The estimates of $\theta_2(\lambda)$ are approximately  
equal to the theoretical value  $\theta_2^{(RG)}=0.71$ for 
$0.1 < \lambda < 0.5$. 
 
We evaluated exponents $\theta(i,j)$ 
which describe the short-distance interchain correlation 
between vertices $i$ and $j$ of $N$-step SAW. They generalize 
des Cloizeaux's three exponents $\theta_s$ for $s=0, 1, 2$.    
Expressing vertices $i$ and $j$ in terms of parameters $\lambda$ and $\mu$ 
we observed the crossover of exponents $\theta(i,j)$: 
from $\theta_2$ to $\theta_1$ 
and from $\theta_1$ to $\theta_0$ as $\lambda$ approaches 1; 
exponent $\theta(i, j)$ changes from $\theta_2$ to $\theta_1$ as 
parameter $\mu$ approaches 0.

We have shown that formula (\ref{eq:universal-f-SAP}) gives 
good fitting curves to the data of the probability distribution functions 
of the distance between two points of cylindrical SAP. 
For the thick cylinder case of cylindrical radius $r_{ex}=0.1$ 
the estimates of exponent $\theta(\lambda)$ for  
the short-distance correlation is a little smaller than 
the theoretical value $\theta_2^{(RG)}$ of SAW.  

Finally, we suggest that the results of this paper 
should be useful for studying the scaling behaviour 
of interchain correlation for toplogical polymers with 
more complex structures.

\section*{Acknowledgement}

The authors would like to thank B. Duplantier for bringing us 
Ref. \cite{desCloizeaux}. They are grateful to Dr. A. Yao 
for helpful comments.

\vskip 24pt


\end{document}